\documentstyle[floats,aps,epsfig,eqsecnum]{revtex}

\begin{document}

\draft
\twocolumn[\hsize\textwidth\columnwidth\hsize\csname@twocolumnfalse\endcsname

\author{Ranjan Mukhopadhyay, C.L. Kane, and T.C. Lubensky}
\address{Department  of Physics, University of Pennsylvania, 
Philadephia, Pennsylvania 19104}
\title{Sliding Luttinger liquid phases}
\date{{\today}}
\maketitle

\begin{abstract}
 We study systems of coupled spin-gapped and gapless Luttinger liquids.
First, we establish the existence of a sliding Luttinger liquid
phase for a system of weakly coupled parallel quantum wires,
with and without disorder. It is shown that the coupling can 
{\it stabilize} a Luttinger liquid phase in the presence of disorder.
We then extend our analysis to a system of crossed Luttinger
liquids and establish the stability of a non-Fermi liquid state:
the crossed sliding Luttinger liquid phase (CSLL). In this
phase the system exhibits a finite-temperature, long-wavelength, isotropic
electric conductivity that diverges as a power law in temperature $T$
as $T \rightarrow 0$. This two-dimensional system has many properties of 
a true isotropic Luttinger liquid, though at zero temperature it
becomes anisotropic. An extension of this model to a three-dimensional
stack exhibits a much higher in-plane conductivity than the
conductivity in a perpendicular direction.
\end{abstract}


\pagebreak 
\vskip2pc]
\narrowtext 

\section{Introduction}

For over two decades a central theme in the study of correlated
electronic systems
has been the drive to understand and classify electronic states that 
do not conform to Landau's Fermi-liquid theory.  A clear example of such
``non-Fermi liquid"
physics occurs in one dimension (1D)\cite{voit}, where arbitrarily weak
interactions destroy the Fermi surface and 
invalidate the notion of independent quasiparticles at low energy.  
Away from charge- and spin-density-wave instabilities, the  1D
interacting electron gas
forms a Luttinger liquid.  The discontinuity in occupation at the Fermi
energy is replaced by
a power-law singularity, and the low-lying excitations are bosonic
collective modes in which spin
and charge decouple.  

In this paper, we investigate how such non-Fermi liquid behavior
could arise in two- and three-dimensional
systems consisting of
arrays of quantum wires or chains. Our analysis
is largely motivated by the unusual normal-state
properties of High $T_{c}$ materials. 
Prominent among these are:
\begin{enumerate}
\item Resistivity perpendicular to the CuO$_{2}$ plane  much larger
than the in-plane resistivity \cite{resistance}.
\item Linear temperature ($T$) dependence of resistivity
along the conducting planes, and $1/T^{2}$ divergence of
the Hall angle \cite{ong}.
\item Angle resolved photoemission (ARPES) data showing a pseudogap
and absence of dispersion in the $c$ direction \cite{ARPES}. 
\item Linear temperature dependence of the Nuclear
Magnetic Resonance (NMR) relaxation $1/T_{1}$ \cite{NMR}. 
\end{enumerate}
Anderson has suggested that these unusual normal-state
properties of the cuprates are the result of 
non-Fermi-liquid physics in two dimensions\cite{anderson1}. 
The study of non-Fermi liquids in  dimensions greater than one 
has, however, proven to be quite difficult.  Since the
Fermi liquid is stable for weak interactions, perturbative methods
about this state fail to locate non-Fermi liquid states\cite{randeria}.  
Moreover, generalizations of the bosonization technique to isotropic
systems in higher dimensions 
have indicated that Fermi-liquid theory survives provided the
interactions are not pathologically 
long-ranged\cite{marston}.  

An alternative approach has been to study
anisotropic systems consisting 
of arrays of parallel weakly coupled 1D wires\cite{coupled}.
Coupled Luttinger liquids have been studied extensively for the past 
thirty years, mostly in the context of
quasi one-dimensional conductors.
Single-particle hoppings as well as pair-hopping
correlations were shown to
destabilize the Luttinger liquid phase. 
Following Anderson's
suggestions regarding high-temperature superconductors,
there have been a series of  studies on coupled
quantum chains, which confirm this view \cite{coupled2} though
the issue remains controversial \cite{anderson2}. It has 
recently been proposed\cite{efkl,vc} 
that for a range of interwire charge and current 
interactions, there is a {\it smectic-metal} (SM) phase in which
Josephson, charge- and spin-density-wave, and single-particle couplings are
{\it irrelevant}.  This phase is a two or three-dimensional
 anisotropic sliding Luttinger liquid  
whose transport properties exhibit power-law singularities like those of a
1D Luttinger liquid.  It is
the quantum analog of the sliding phases of coupled classical XY
models first identified in the context of
phases of DNA-lipid complexes  \cite{sliding,olt}.
The study of coupled Luttinger liquids becomes particularly
relevant in the context of striped phases which have  
recently been found in quantum Hall systems \cite{oppen} and 
in the cuprates \cite{tranquada}.
These phases are characterized by inhomogenious 
distributions of charge and spin, where, for example,
the charge carriers might be confined to separate
linear regions, thus resembling stripes.  Sliding
Luttinger Liquid phases might also be relevant for a variety of
other systems such as quasi one-dimensional organic
conductors and ropes of nanotubes.

   The work on sliding Luttinger liquid phases has also been extended to
a square network of 1D wires formed by
coupling two perpendicular smectic metals \cite{mlk},
which has been shown to exhibit a
{\it crossed sliding Luttinger Liquid} (CSLL) phase. 
At finite temperature $T$, the CSLL phase is a
$2D$ Luttinger liquid with an isotropic long-wavelength
conductivity that diverges as a power-law in $T$ as $T\rightarrow 0$.  At
$T=0$, it is essentially two independent smectic metals.
This model could be realized in man-made structures
constructed from quantum wires such as carbon nanotubes.
Extension of the model to a three-dimensional 
stack may be relevant to the stripe phases of the cuprates.
Based on neutron and x-ray scattering measurements, 
it has been 
suggested that spin-charge stripes in the adjacent CuO$_{2}$ plane
are orthogonal to each other
\cite{tranquada}.         

   In order to understand the nature of the sliding phase, it
is easiest to think in terms of the classical analog.
Imagine a stack of 2D XY models coupled by the Josephson
coupling $\cos(\theta_{n}-\theta_{n+1})$ where $\theta_{n}$
is the XY-angle variable in layer $n$. Such a system
always goes directly from a 3D ordered phase to a completely disordered
phase as a function of the temperature $T$. However 
in the presence of interlayer gradient
coupling terms of the form $\nabla \theta_{m} \cdot \nabla \theta_{n}$,
the system may have an intermediate sliding phase
\cite{olt}, that
exhibits an in-plane 2D order with power-law decay of correlations
along the planes. In the absence of Josephson couplings, in 
the sliding phase, $\theta$ in neighboring layers
freely  slide over each other with no energy cost, and two-point
correlation functions are identical in form to those of a stack
of decoupled 2D layers. When inter-layer Josephson couplings
are present, though irrelevant, two-point correlation
functions decay exponentially perpendicular to the layers.
Thus the sliding phase has in-plane 2D order but is 
disordered in the third direction. For sliding Luttinger
liquid models, the 2D layers are replaced by 1D quantum 
wires, and inter-wire current-current and density-density
couplings play the role of the gradient couplings. Thus, in
the sliding Luttinger liquid phases, the correlation functions along
a given wire exhibit the same power-law functional form as
in a 1D Luttinger liquid.

    In this paper, we review and discuss in detail 
results on the sliding Luttinger liquid phase. Our discussion
of the parallel Luttinger liquids follows Ref.~\cite{vc}, but
it establishes the stability of the sliding phase to a more
complete set of inter-wire operators and also to disorder.
For the crossed sliding Luttinger liquid phase (CSLL), we provide
detailed calculations that were presented only briefly in
Ref.~\cite{mlk}. The stability and transport properties
of the CSLL phase form the central results of the paper.
We also extend the analysis to a three-dimensional stack  
of crossed Luttinger liquids, and show that we can still
obtain a sliding Luttinger liquid phase, with hopping between
planes being irrelevant. In this phase, the in-plane
and perpendicular conductivity can be quite different and exhibit
different temperature dependences. This could be
of relevance to the normal state of the cuprates.

     This paper is organized as follows. In Sec.~II we 
explore the stability of the 
sliding phase for an array of parallel coupled Luttinger liquids
with respect to a large (but not complete) set of inter-wire operators.
For a one-dimensional system of interacting spin-$1/2$ fermions, 
the spin excitations could either be gapped or
gapless. In the spin-gapped Luther-Emery regime, the system can by
described by a single Luttinger liquid for charge. In the gapless
case, both spin and charge are dynamical degrees of freedom, and there
are two Luttinger parameters ($\kappa_{\sigma}$, $\kappa_{\rho}$), and 
two velocities ($v_{\sigma}$, $v_{\rho}$). We study 
coupled Luttinger liquids both for the spin-gapped and the gapless case,
and in each case demonstrate the stability of the sliding
phase. In addition, we study the effect of disorder. We find
that density-density and current-current interactions that
stabilize the sliding phase, also make disorder more 
strongly irrelevant. Note, however, that in this paper
we do not establish stability with respect to all 
multi-wire operators. Since the overall strength of
these higher order interactions that we neglect
are expected to be much smaller than 
those we consider here, their relevance becomes important
only at very small temperatures. 
We delay a more complete study of their effects to
a future publication. Throughout the paper we will use
the term ``stable'' somewhat loosely to refer to stability with
respect to a restricted set of interactions that include
all two-wire interactions. In Sec.~III we
extend the above analysis to two identical coupled
Luttinger liquid arrays, arranged such that the
wires of one array run perpendicular to those of the other.
In this case we establish the stability of a crossed sliding Luttinger
liquid (CSLL) phase to a large class of operators. The inter-array
density-density couplings effectively renormalize the
intra-array couplings; however, the stability of the CSLL
phase turns out to be identical to the stability of the 
smectic-metal phase for each array, but with the intra-array
couplings replaced by the renormalized couplings.
 We then generalize our analysis
to a three-dimensional stack of arrays, with wires
on each array running perpendicular to the wires
of the consecutive array, and obtain once more
a sliding phase that appears stable.
In Sec~IV we explore the transport properties of sliding phases.
In particular, we focus on the power-law singularities of the conductivity
as a function of $T$ as $T \rightarrow 0$.
Finally Sec~V sums up our principal results. Appendix~A sketches 
out the steps used to obtain intra-wire correlation functions,
Appendix~B 
and Appendix~C present details 
of some integrals used in Secs.~II and III, and Appendix~D presents details
of conductivity calculations using the Kubo formula.

\section{Coupled parallel Luttinger liquids}
     Our goal is to construct non-Fermi-liquid electron systems
in two and three dimensions. Our basic building blocks are 
one-dimensional quantum wires that exhibit Luttinger-liquid
phases with non-Fermi liquid power-law decay of correlation
functions. We couple these wires together in arrays
in such a way that they retain their one-dimensional 
non-Fermi liquid character yet allow non-vanishing interwire
electron transport at nonzero frequency or temperature. We will consider
several types of arrays, and it is useful to introduce a nomenclature 
for them. The long axis of parallel wires defines one direction in space.
The wires can be arranged in either a one- or a two-dimensional
Bravais lattice of points in the plane perpendicular to that direction.
We, therefore introduce the notation
$d_{\perp}$:1 to denote an array of parallel wires centered on
lattice points in a $d_{\perp}$-dimensional space.
The resulting structure occupies a $d_{\perp}+1$ dimensional
space. Thus a 1:1 array is a planar array of equally spaced
parallel wires, and a 2:1 array is a three-dimensional
columnar array of wires. As stated, we will restrict our attention
to arrays in which the wires lie on a $d_{\perp}$-dimensional periodic 
lattice. For 2:1 arrays, we will only consider, in detail, 
those based on a 2-dimensional rectangular lattice.
Both 1:1 and 2:1 arrays can exhibit anisotropic sliding 
phases for an appropriate choice of interwire potentials.

     In this section we will consider anisotropic sliding 
metal phases in 1:1 and 2:1 arrays. In the succeeding sections,
we will discuss how crossed 1:1 and 2:1 arrays can produce two and
three-dimensional sliding phases with $C_{4v}$ symmetry 
and isotropic long wavelength finite temperature 
conductivity. We begin our discussion with a review of
1D Luttinger liquids.

\subsection {Review of 1D Luttinger liquids}
One dimensional Luttinger liquids are most easily
described in terms of collective  bosonic charge and spin-density 
modes.  The bosonized form of the fermion operators
near the left and right Fermi points are \cite{delft}
\begin{eqnarray}
{R}_{s,j}(x)&=& {1 \over \sqrt{4 \pi \epsilon}} {\eta}_{R,s,j}
		  e^{i k_{F}x}  e^{i \Phi_{R,s,j}(x)} \nonumber \\
{L}_{s,j}(x)&=& {1 \over \sqrt{4 \pi \epsilon}} { \eta}_{L,s,j}
		e^{- i k_{F}x}  e^{i \Phi_{L,s,j}(x)},
\label{fermions}
\end{eqnarray}
where $R$ and $L$ stand for the right and left moving
electrons, respectively, $s$ is the spin index, $j$ denotes
the wire number, $\epsilon$ is some intra-chain cutoff, and $\eta_{R/L,s,j}$
are the Klein factors. We can write down an effective theory
for the low-energy excitations in terms of the boson operators $\Phi$.
It is convenient to define
\begin{eqnarray}
\theta_{s,j}&=&(\Phi_{R,s,j}+\Phi_{L,s,j})/\sqrt{4 \pi}, \nonumber \\
\phi_{s,j}&=&(\Phi_{R,s,j}-\Phi_{L,s,j})/\sqrt{4 \pi},
\end{eqnarray}
where $\theta_{s,j}$ is a phase variable, and $\phi_{s,j}$ is the
conjugate density variable \cite{notation}. Since the bosonic excitations
can be characterized in terms of charge and spin excitations, 
we further define
\begin{equation}
   \theta_{\rho,j}=(\theta_{\uparrow,j}+\theta_{\downarrow,j})/\sqrt{2},
\ \ \ \theta_{\sigma,j}=(\theta_{\uparrow,j}-\theta_{\downarrow,j})/\sqrt{2},
\end{equation}
and similarly for the $\phi$ variables. Here $\rho$ characterizes
the charge excitations, and $\sigma$ the spin excitations.
For a single Luttinger liquid, we can write down an effective 
Hamiltonian for the charge sector
\begin{equation}
  H_{\rho} =\int dx {v_{\rho} \over 2}\left[{{(\partial_{x}\theta_{\rho})^{2}
} \over \kappa_{\rho}} + \kappa_{\rho}(\partial_{x}\phi_{\rho})^{2}
\right],
\label{ll}
\end{equation}
where $v_{\rho}$ is the velocity, and $\kappa_{\rho}$ is a 
Luttinger liquid parameter (this is the inverse of the
usual Luttinger liquid parameter $g$).
Alternately, we could write down the action, in a
path-integral formulation, as
\begin{eqnarray}
  S_{\rho} &=&\int dx d\tau \big\{ {v_{\rho} \over 2}
\left[{{(\partial_{x}\theta_{\rho})^{2}
} \over \kappa_{\rho}} + \kappa_{\rho}(\partial_{x}
\phi_{\rho})^{2}\right]\nonumber \\ 
& & - 2 i (\partial_{x}\theta_{\rho})(\partial_{\tau}\phi_{\rho})\big\}.
\end{eqnarray}
Integrating out either the $\theta_{\rho}$ or the $\phi_{\rho}$ variables,
we obtain
\begin{eqnarray}
S_{\rho,\phi} &=&{\kappa_{\rho} \over 2} \int dx d\tau \left[v_{\rho} (\partial_{x} 
\phi_{\rho})^{2}
+ {1 \over v_{\rho}} (\partial_{\tau} \phi_{\rho})^{2} \right] \nonumber \\
S_{\rho,\theta}&=&{1 \over {2 \kappa_{\rho}}} \int dx d\tau \left[
v_{\rho} (\partial_{x} 
\theta_{\rho})^{2}
+ {1 \over v_{\rho}} (\partial_{\tau} \theta_{\rho})^{2} \right]. 
\end{eqnarray}
In the spin sector, we might either have a spin gap corresponding
to the Luther-Emery regime, or have a gapless
phase where the spin excitations are described by
a Hamiltonian of the same form as Eq.~(\ref{ll})
with the parameters $\kappa_{\sigma}$ and $v_{\sigma}$.
In the gapless phase, SU(2) symmetry imposes the
constraint $\kappa_{\sigma}=1$. 

\subsection{The spin-gapped 1:1 array}

           In this subsection we consider the simplest
array of quantum wires, the two-dimensional 1:1 array
of Luttinger liquids in the spin-gapped phase.
It has been suggested that this
case might describe the striped phases of
high-temperature superconductors \cite{emery}.
In this subsection all bosonic variables refer to the
charge sector, and we do not write explicitly the subscript
$\rho$. In general, we expect density-density 
and current-current interactions between the wires, 
which can be represented by a action of the form 
\begin{equation}
S_{\rm int}={1 \over 2} \sum_{n,n',\mu} \int dx d\tau
j_{\mu,n}(x,\tau) \tilde{W}_{\mu}(n-n')j_{\mu,n'}(x,\tau) ,
\label{lagrangian2}
\end{equation}
where $j_{\mu ,n}=(\rho_{n}(x,\tau),J_{n}(x,\tau))$ with
$\rho_{n}=\partial_{x}\phi_{,n}(x,\tau)$ the density and
$J_{n}=\partial_{x}\theta_{n}(x,\tau)$ the current on the $n$th
wire. 
The density-density interaction is  an  effective interaction
generated by both the screened Coulomb
and  the electron-phonon interaction.
In the striped phases,  stripe fluctuations lead to
 current-current interactions \cite{emery}.
These current-current and density-density interactions are marginal and should be included in the
fixed-point action. They are invariant under the ``sliding"
transformations $\phi_{n} \rightarrow \phi_{n} + \alpha_{n}$ and
$\theta_{n} \rightarrow \theta_{n} + \alpha^{\prime}_{n}$. Equations
(\ref{ll}) and (\ref{lagrangian2}) define the fixed-point
action of the smectic-metal phase\cite{efkl}, which can be written
as
\begin{eqnarray}
  S&=&\sum_{n}\int dx d\tau [\sum_{j} V^{\theta}_{j}
(\partial_{x}\theta_{n})(\partial_{x}\theta_{n+j})
\nonumber \\
 &+& \sum_{j} V_{j}^{\phi}(\partial_{x}
\phi_{n})(\partial_{x} \phi_{n+j}) 
+ 2 i (\partial_{x}\theta_{n})(\partial_{\tau}\phi_{n})].
\end{eqnarray}
Upon integration over $\phi_{n}$ or $\theta_{n}$, respectively,
the effective action for $\theta_{n}$ and $\phi_{n}$ become
\begin{eqnarray}
S_{\theta} &=& \int {{d^{3}Q}
\over {(2 \pi)^{3}}} 
{1 \over 2}\kappa(q_{\perp})\left\{ {1 \over v({q}_{\perp})}
\omega^{2} + v({\vec q}_{\perp})q_{\parallel}^{2}\right\}  |\phi({\bf Q})|^{2}
 \nonumber \\
S_{\phi}&=&\int {{d^{3}Q}
\over {(2 \pi)^{3}}} 
{1 \over 2\kappa(q_{\perp})}\left\{ {1 \over v({q}_{\perp})}
\omega^{2} + v({\vec q}_{\perp})q_{\parallel}^{2}\right\}
 |\theta({\bf Q})|^{2}
\end{eqnarray}
where ${\bf Q}=(\omega, q_{\parallel}, {q}_{\perp})$, with $q_{\parallel}$
the momentum along the chain and ${q}_{\perp}$ that perpendicular to
the chains. Here 
\begin{eqnarray}
\kappa(q_{\perp})&=&\sqrt{V^{\phi}(q_{\perp})/ 
V^{\theta}(q_{\perp})} \nonumber \\
v(q_{\perp})&=&\sqrt{V^{\phi}(q_{\perp}) V^{\theta}(q_{\perp})},
\end{eqnarray}
where $V^{\theta}(q_{\perp})$ and $V^{\phi}(q_{\perp})$
are the Fourier transforms of $V^{\theta}_{j}$
and $V^{\phi}_{j}$ with respect to the wire index $j$.
Eqs.~(2.8) and (2.9) define the action of the ideal 
2D sliding Luttinger liquid or SM phase. Even though 
they include interwire interactions described by Eq. (2.7),
they yield power-law correlations characteristic of a 1D
Luttinger liquid.

   We could consider, for example, correlation functions
involving the density variable $\rho$. We note that 
\begin{eqnarray}
\rho\equiv\psi^{\dagger}\psi &=&R^{\dagger} R
+ L^{\dagger} L  + [L^{\dagger} R + {\rm c.c.}] \\ 
\nonumber
&\simeq& \partial_{x} \phi + [e^{i 2 k_{F} x + i \sqrt{2 \pi} \phi(x, \tau)}
+ {\rm c.c.}] .
\end{eqnarray}
Thus the density has two pieces: the first piece is the course-grained
density and the second is modulated at $2 k_{F}$ where $k_{F}$ is
the Fermi wave-vector. Correspondingly, the 
density-density correlation has two pieces. We consider
 $G_{\phi}(x,\tau)$: the component of the density-density correlation function
with $2 k_{F}$ modulation. Thus 
$$G_{\phi}(x,\tau) \equiv \langle
e^{i \sqrt{2 \pi}(\phi_{j}(x,\tau) -  \phi_{j}(0,0)) + i 2 k_{F} x} \rangle
+ {\rm c.c.}. $$
It easy to show that for large $x$ (see Appendix A),
\begin{equation}
 G_{\phi}(x,0) \approx {{A_{1} \cos(2 k_{F} x)}
\over { x^{ \Delta_{{\rm CDW},\infty}}}},
\end{equation}  
where
\begin{equation}
\Delta_{{\rm CDW},\infty} = \int_{-\pi}^{\pi} {{dq_{\perp}} \over 2\pi}
{1 \over {\kappa(q_{\perp})}},
\end{equation}
and $A_{1}$ is a constant. Alternately, for large $\tau$
\begin{equation}
 G_{\phi}(0,\tau) \approx A_{2} \tau^{- \Delta_{{\rm CDW},\infty}}
\end{equation}
where $A_{2}$ is a different constant. In general, $G_{\phi}(x,\tau)$
can be written in the scaling form
\begin{equation}
 G_{\phi}(x,\tau) \approx x^{- \Delta_{{\rm CDW},\infty}} F\left(
{x \over \tau}\right)
\end{equation}
where
\begin{eqnarray}
F(y) &\rightarrow& A_{1} \ \ \ \ \ \ \ \
 {\mathrm as } \ \  y \rightarrow 0 \nonumber \\
     &\rightarrow& A_{2} y^{\Delta_{{\rm CDW},\infty}} \ \
{\mathrm as } \ \  y \rightarrow \infty. 
\end{eqnarray}

\begin{figure}
\par\columnwidth20.5pc
\hsize\columnwidth\global\linewidth\columnwidth
\displaywidth\columnwidth
\epsfxsize=2.0truein
\centerline{\epsfbox{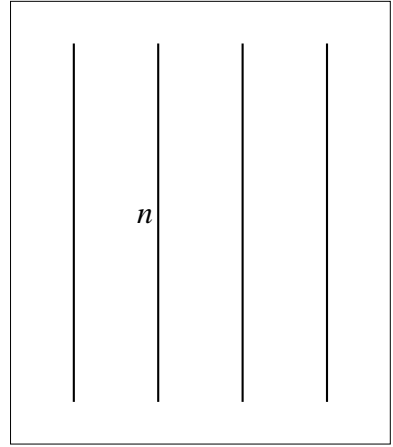}}
\vspace*{.1in}
\caption{A schematic depiction of a 2-dimensional array of quantum
wires} 
\label{fig:wires1}
\end{figure}
      
A variety of interactions, other than those of Eq.~(2.7) couple
neighboring wires. The sliding phase is stable only if the 
interactions are irrelevant, that is, only if they scale to
zero with system size. We will now investigate perturbatively
the relevance of these interchain
 interactions to determine under what conditions
the sliding phase is stable. Due to the spin
gap, single-particle hopping between chains is irrelevant, and the
inter-chain interactions that could become relevant are 
the Josephson (SC) and CDW couplings, which are represented by operators
of the form
\begin{eqnarray}
   {\cal H}_{{\rm SC},n}&=& \sum_{j} \int dx [
(R^{\dagger}_{\uparrow,j} {L}^{\dagger}_{\downarrow,j}
+{R}^{\dagger}_{\downarrow,j} {L}^{\dagger}_{\uparrow,j})
\nonumber \\
& &\times ({R}_{\uparrow,j+n} {L}_{\downarrow,j+n}
+ {R}_{\downarrow,j+n} {L}_{\uparrow,j+n})] + {\rm c.c.},
\end{eqnarray}
and
\begin{eqnarray}
   {\cal H}_{{\rm CDW},n}&=& \sum_{j} \int dx[
({R}^{\dagger}_{\uparrow,j} {L}_{\uparrow,j}
+{R}^{\dagger}_{\downarrow,j} {L}_{\downarrow,j})
\nonumber \\
& &\times ({L}^{\dagger}_{\uparrow,j+n} {R}_{\uparrow,j+n}
+ {L}^{\dagger}_{\downarrow,j+n} {R}_{\uparrow,j+n})] + {\rm c.c.}.
\end{eqnarray}
At low temperatures the spin variable $\phi_{\sigma}$ is
effectively frozen, and these interactions depend only
on $\theta_{i}$. Their associated actions  can be expressed as:
\begin{eqnarray}
S_{{\mathrm SC},n} &=& {\mathcal{J}}_{n} \sum_{i} \int dx d\tau  \cos[\sqrt{2
\pi}(\theta_{i} - \theta_{i+n})], \nonumber \\ 
S_{{\mathrm CDW},n} &=&
{\mathcal{V}}_{n}\sum_{i}\int dx d\tau  \cos[\sqrt{2 \pi}
(\phi_{i} - \phi_{i+n})].
\label{interactions1}
\end{eqnarray}
The relevance of these terms are determined by the scaling
dimensions of the corresponding operators,
$\cos[\sqrt{2 \pi}(\theta_{i} - \theta_{i+n})]$
and $\cos[\sqrt{2 \pi}(\phi_{i} - \phi_{i+n})]$,
which are, respectively,
\begin{eqnarray}
\Delta_{{\mathrm SC},n}&=&\int_{- \pi}^{\pi} {{d q_{\perp}} \over {2 \pi}}
(1 - \cos{n q_{\perp}}) \kappa(q_{\perp}) \nonumber \\
\Delta_{{\mathrm CDW},n}&=&\int_{- \pi}^{\pi} {{d q_{\perp}} \over {2
\pi}} {{(1 - \cos{n q_{\perp}})} \over {\kappa(q_{\perp})}}.
\label{stabilitysc}
\end{eqnarray}
 The exponent $\Delta_{{\rm SC},n}$ follows from
\begin{equation}
\langle \cos\sqrt{2 \pi}(\theta_{i} - \theta_{i+n})\rangle
= e^{- \pi \langle (\theta_{i} - \theta_{i+n})^{2} \rangle},
\end{equation}
and
\begin{eqnarray}
 {1 \over 2}\langle (\theta_{i} - \theta_{i+n})^{2}\rangle
&=& \int {{d^{3}Q} \over {(2 \pi)^{3}}}
  \langle|{\tilde \theta}({\bf q}, \omega)|^{2}\rangle(1 - \cos(q_{\perp} n))
\nonumber \\
&=&\int_{-\pi}^{\pi}{d q_{\perp} \over 2\pi} \{\kappa(q_{\perp})
(1 - \cos q_{\perp}n) \nonumber \\
& &\times \left[\int {{dq_{\parallel}d\omega} \over {(2 \pi)^{2}}}
{1 \over {v q_{\parallel}^{2} + \omega^{2}/v}}\right] \} .
\end{eqnarray}
where $\langle \rangle$ denotes averaging with respect to $S$
in Eq.~(2.8). 
The integral in the square brackets diverges logarithmically with
system size $L$ ($\sim C \ln L$) \cite{footnote1}.
Using this, we find
\begin{equation}
\langle \cos\sqrt{2 \pi}(\theta_{j} - \theta_{j+n})\rangle
\sim L^{-\Delta_{{\rm SC},n}},
\end{equation}
where $\Delta_{{\rm SC},n}$ is given by Eq.~(\ref{stabilitysc}).
From the above equation, it follows that
\begin{equation}
\langle S_{{\rm SC},n} \rangle \sim L^{2 - \Delta_{{\rm SC},n}}.
\end{equation} 
Similar calculations produce $\Delta_{{\rm CDW},\infty}$.
For a stable smectic metal phase, these terms have to be
irrelevant, implying
\begin{equation}
\Delta_{{\mathrm CDW},n} > 2, \ \ \  \Delta_{{\mathrm SC},n^{'}} > 2
\label{irrelevance}
\end{equation}
for all $n$ and $n^{'}$.

\begin{figure}
\par\columnwidth20.5pc
\hsize\columnwidth\global\linewidth\columnwidth
\displaywidth\columnwidth
\epsfxsize=3.0truein
\centerline{\epsfbox{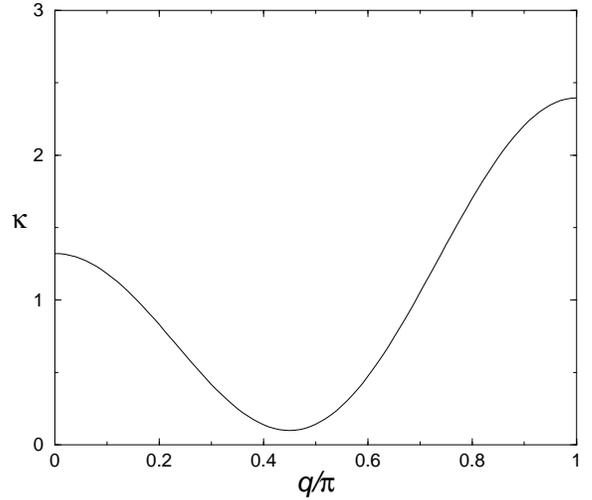}}
\vspace*{.1in}
\caption{$\kappa(q)$ as a function  of $q/\pi$, with a minimum,
$\kappa_{\rm min}$ at $q=q_{0}$.}
\label{fig:kappa1}
\end{figure}

    If any $\Delta_{{\rm SC},n} < 2$, the smectic metal (SM)
phase is unstable to the formation of an anisotropic 2D superconductor. If any
$\Delta_{{\rm CDW},n}<2$, the SM phase will flow to a 2D 
longitudinal CDW-crystalline phase with $2 k_{F}$ density
modulations along the wires and phase locked from wire to wire.
Notice that if $\kappa(q_{\perp})$ is uniformly small, $\Delta_{{\rm SC},n}$'s
are small and  $\Delta_{{\rm CDW},n}$'s are large, and for large $\kappa$,
$\Delta_{{\rm SC},n}$'s
are large and  $\Delta_{{\rm CDW},n}$'s small.  For a stable sliding
phase we need all $\Delta_{{\rm SC},n}$'s
and $\Delta_{{\rm CDW},n}$'s to be greater than two. Our strategy
to create a stable sliding phase
is to choose $\kappa(q_{\perp})$ of the form shown in Fig.~(2),
with $\kappa$ having a minimum, $\kappa_{\rm min}$, at $q_{\perp}
= q_{0}$. When $\kappa_{\rm min}$ becomes zero, the system
undergoes a transition to a transverse CDW modulation
with wavevector $q_{0}$,
where the charge-density varies from wire to wire.
For small but positive $\kappa_{\rm min}$, the system is
close to the transverse CDW instability. As pointed out in
\cite{vc}, a transverse CDW would frustrate crystallization of 
fermions, since $k_{F}$ is now a function of the chain index $j$.
Strong fluctuations of this kind  prevent the locking in of 
density fluctuations along the wires. Thus, in order to 
stabilize the sliding Luttinger liquid phase, we need 
to tune $\kappa_{\rm min}$ to be very small compared
to average $\kappa(q_{\perp})$, so that 
$\Delta_{{\rm SC},n}$
and $\Delta_{{\rm CDW},n}$ can both be made large.
Note that in addition we should also consider
interchain operators 
of the form ${R}^{\dagger}_{\uparrow,j}{R}^{\dagger}
_{\downarrow,j}{R}_{\uparrow,j+n}{R}_{\downarrow,j+n}$.
These interactions, however, turn out to be automatically irrelevant
if the superconducting and CDW interactions are irrelevant, and hence 
merit no further consideration. 

    In addition to the lowest-order interactions between 
pairs of chains described by $S_{{\rm SC},n}$ and $S_{{\rm CDW},n}$,
there can in principle be higher-order multi-chain interactions
with actions of the form 
\begin{eqnarray}
{S}_{{\mathrm SC},s_{p}} &=& \sum_{i} \int dx d\tau {\mathcal{J}}_{s_{p}} 
\cos[\sqrt{2
\pi}(\sum_{p} s_{p}\theta_{i+p})], \nonumber \\ 
{S}_{{\mathrm CDW},s_{p}} &=& \sum_{i} \int dx d\tau
{\mathcal{V}}_{s_{p}} \cos[\sqrt{2 \pi}(\sum_{p}s_{p}
\phi_{i+p})]
\end{eqnarray}
where ${\mathcal{J}}_{s_{p}}$ are the inter-chain Josephson couplings,
${\mathcal{V}}_{s_{p}}$ the inter-chain particle-hole (CDW) interactions,
and $s_{p}$ is an integer-valued function of the chain number $p$
satisfying $\sum_{p} s_{p}=0$.
The scaling dimensions of 
$\cos[\sqrt{2 \pi}(\sum_{p} s_{p}\theta_{i+p})]$ and
$\cos[\sqrt{2 \pi}(\sum_{p} s_{p}\phi_{i+p})]$ 
are, respectively, 
\begin{eqnarray}
\Delta_{{\mathrm SC},s_{p}}&=& \int_{- \pi}^{\pi} {{d q_{\perp}} \over {2 \pi}}
 \kappa(q_{\perp})\left[\sum_{p,p^{'}}
    s_{p}s_{p^{'}}\cos\{(p - p^{'})q_{\perp}\}\right], \nonumber  \\
\Delta_{{\mathrm CDW},s_{p}}&=&\int_{- \pi}^{\pi} {{d q_{\perp}} \over {2
\pi}} {1 \over {\kappa(q_{\perp})}}\left[\sum_{p,p^{'}}
    s_{p}s_{p^{'}}\cos\{(p - p^{'})q_{\perp}\}\right]. \nonumber \\
\label{cdwsc}
\end{eqnarray}
These perturbations are irrelevant if 
\begin{equation}
\Delta_{{\mathrm CDW},s_{p}} > 2, \ \ \  \Delta_{{\mathrm SC},s^{'}_{p}} > 2
\label{stability}
\end{equation}
for all sets of $s_{p}$ and $s^{'}_{p}$. The relevance of
higher order terms of this form is a subtle issue,  and we will
elaborate on this in a future publication.
However the strength of these terms,
as measured for example by ${\mathcal{V}}_{s_{p}}$, 
is small and they become important only at very small
temperatures even if they are relevant.
In this paper we will restrict
ourselves to pairwise CDW couplings of the form given by Eq.~(2.17);
 however,
we  comment on the higher order superconducting terms below.

To explore the regions of stability of the smectic metal
(SM) phase, we follow
Refs. \cite{vc} and \cite{olt} and take
\begin{equation}
\kappa(q_{\perp})= K[1 + \lambda_{1}\cos(q_{\perp}) + \lambda_{2}
\cos(2 q_{\perp})] .
\label{kappa}
\end{equation}
The parameters $\lambda_{1}$ and $\lambda_{2}$ can be tuned to set the value 
$q_{0}$ of $q_{\perp}$ at which $\kappa(q_{\perp})$ reaches a minimum
and the minimum value of $\kappa(q_{\perp})$, $\kappa(q_{0})=K\Delta$.
For a specific value of $\Delta$ and $k$,
\begin{eqnarray}
   \lambda_{1}&=& - {{4 (1 - \Delta) \cos q_{0}} \over
	        (1 + 2 \cos^{2}(q_{0}))}  \nonumber \\
   \lambda_{2}&=& {(1 - \Delta) \over (1 + 2 \cos^{2} q_{0})},
\label{lambdas}
\end{eqnarray}
unless $q_{0}=0$ or $\pi$, in which case only $\lambda_{1}
+ \lambda_{2}$ is fixed by $\Delta$. We note that in the  
above equation (i.e. for $q_{0} \neq 0,\pi$), while $\lambda_{2}$
is always positive for $\Delta<1$, $\lambda_{1}$ can be 
either positive or negative.
Typically positive $\lambda$'s correspond to repulsive interactions.

                                      In this paper we
treat $K$ as the control parameter. For small $K$
the system becomes superconducting while for large $K$ the system
goes into a CDW crystalline phase. We will show that through
judicious tuning of $\lambda_{1}$ and $\lambda_{2}$ it
is possible to have an intermediate window of $K$ where
the smectic metal phase is stable. For this purpose, we define  
$a_{s_n}=\Delta_{{\mathrm SC},s_{n}}/ K$ and 
$b_{n} = \Delta_{{\mathrm CDW},n}K$, where 
$a_{s_n}$ and $b_{n}$ depend only on $\lambda_{1}$
and $\lambda_{2}$, and  
\begin{eqnarray}
a_{s_p}&=&\sum_{p}{1 \over 2}(s_{p}^{2}+
s_{p}(s_{p+1}+s_{p-1})\lambda_{1}/2 \nonumber \\
& & + s_{p}(s_{p+2} + s_{p-2})
\lambda_{2}/2).
\end{eqnarray} 
The SM phase becomes unstable
to inter-chain Josephson couplings for $K$ less than
$K_{\mathrm{SC}}={\mathrm{max}}_{s_{p}} (2/a_{s_{p}})$ and unstable to 
inter-chain CDW
interactions for $K$ greater than $K_{\mathrm{CDW}}={\mathrm
{min}}_{n} (b_{n}/2)$.
Thus the smectic metal phase is stable over a window of $K$,
$K_{\mathrm {SC}}< K < K_{\mathrm CDW}$, provided
\begin{equation}
\beta \equiv {K_{\mathrm{CDW}} \over K_{\mathrm {SC}}}=
 \left. {{a_{s_{p}} b_{m}} \over 4} \right|_
{{\mathrm{min. wrt.}} m \& {s_{p}}}> 1.
\label{beta}
\end{equation}
We note, once more, that in this paper we consider stability
with respect to all superconducting terms, but only pairwise
CDW terms of the form given by (2.17).
If $\beta < 1$, the system goes directly from a 2D superconducting
(SC) phase to a CDW crystal as $K$ is increased, without passing
through the smectic metal (SM) phase. For a stable sliding phase,
we need to make $\Delta$ very small.
The value of $\Delta_{{\rm CDW},n}$
for $\Delta$ small is determined by values of $q_{\perp}$ near
$q_{0}$. We can therefore set $\kappa(q_{\perp})$
\begin{equation}
\kappa(q_{\perp}) \approx K[\Delta + C (q_{\perp} - q_{0})^{2}],
\label{kappaapprox}
\end{equation} 
where 
\begin{equation}
C \equiv {{\kappa''(q_{0})}\over {2 K}} = 2 \lambda_{2} \sin^{2} q_{0}.
\label{C}
\end{equation}
This gives us
\begin{equation}
\Delta_{{\rm CDW},n} \simeq {{K(1 - \cos(n q_{0})e^{-n \sqrt{\Delta/C}})} 
\over \sqrt{C \Delta}},
\end{equation}
where $C$ has been defined in Eq.~(\ref{C}).
We set $\Delta = 10^{-5}$. We consider the range of $q_{0}/\pi$
lying between 0.25 and 0.75. This range can be broken into three
sections:
\begin{enumerate}
\item $0.25 < q_{0}/\pi < 0.41957$. \\
In this range 
$\lambda_{1}<0$ and $|\lambda_{1}|>\lambda_{2}$. Here the
most relevant superconducting term corresponds to the
multi-chain operator $\cos[\sqrt{2\pi}(\theta_{i} + \theta_{i+1}
-\theta_{i+3}-\theta_{i+4})]$ \cite{olt2}. The dimension of this
operators sets the minimum of $a_{s_{p}}$. Thus, in
this range,
${\mathrm min} a_{s_{p}}=(2+\lambda_{1}-\lambda_{2}/2)$
and $K_{\rm SC}= 1/(2+\lambda_{1}-\lambda_{2}/2)$.
\item $0.41957 < q_{0}/\pi < 0.5804 $. \\ 
In this region $|\lambda_{1}|<\lambda_{2}$. We find that
$a_{s_{p}}$ is smallest for the set $s_{n}=\delta_{n,0}-\delta_{n,2}$.
Thus, in this range $K_{\rm SC}=2/(1-\lambda_{2}/2)$.
\item $0.5804 < q_{0}/\pi < 0.75$. \\
Here $\lambda_{1}>\lambda_{2}$, and $a_{s_{p}}$ is the smallest
for the set $s_{n}=\delta_{n,0}-\delta_{n,1}$. Thus 
 $K_{\rm SC}=2/(1-\lambda_{1}/2)$.
\end{enumerate}
In Fig~3 we plot $\beta$ as a function of $q_{0}$. The minima of
the curve correspond to $q_{0}= 2 \pi l/m$, where $l$ and
$m$ are integers. Also note that since  $\lambda_{1}$ has the same sign
as $(-\cos q_{0})$, there are regions of stable smectic
phase for positive as well as negative values of $\lambda_{1}$.

        Having established a stable smectic phase for the
pure system, we now study the relevance of quenched disorder in this
phase. Disorder gives rise to a random electron potential, 
$D(x)$, with associated action
\begin{equation}
{S}_{\rm dis}= \sum_{j} \int dx d\tau D_{j}(x) \cos[\sqrt{2 \pi} 
\phi_{j}].
\end{equation}
$D(x)$ can be treated as a Gaussian random variable,
with zero mean and local fluctuations such that
\begin{eqnarray}
{\overline{D(x)}} &=& 0 \nonumber \\
{\overline {D(x) D(x^{'})}} &=& \Delta_{D} \delta(x-x^{'}). 
\end{eqnarray}
where the over-line signifies averages over the randomness.
By a generalization of the Harris criterion \cite {harris}, it can
be shown quite easily that
(also, see Giamarchi and Schulz \cite{lp}) disorder is irrelevant if
$ \Delta_{{\rm CDW}, \infty}>3$, where
\begin{equation}
\Delta_{\mathrm{CDW},\infty} \equiv \int^{\pi}_{-\pi} {dq \over 2 \pi} {1
\over \kappa(q)} \simeq {1 \over K} {1 \over \sqrt{C \Delta}}.
\end{equation}
For the range of parameters we are considering where the SM phase
is stable, $\Delta_{{\rm CDW}, \infty}$ is large and thus disorder
is strongly irrelevant. This is an important point. For a single
Luttinger liquid in the repulsive region, $\kappa > 1$ and  
disorder is always relevant. However
inter-wire interactions can drive disorder irrelevant for
$\kappa(q_{\perp}=0)>1$, even in regions of phase space where all interactions are
repulsive.  

    Thus, there is a small but finite
region of phase space where the smectic metal phase appears
stable. We should note that over a larger region of phase space
the only relevant operators involve nonlocal interactions
of the form ${\mathcal V}_{n} \cos[\sqrt{2 \pi} 
(\phi_{i} - \phi_{i+n})]$, where ${\mathcal V}_{n}$
is expected to be exponentially small, for large $n$, in the
bare Hamiltonian. Though relevant, these operators would only
play a role for $k_{B} T$ smaller than some energy scale 
set by ${\mathcal V}_{n}$. So, for example, there will be a
range of temperatures where we will only need to consider
the relevance of ${\mathcal V}_{1}$ and ${\mathcal J}_{1}$. 
These can be made irrelevant over a reasonably large
region of phase space (see \cite{efkl}).
Thus, even though the region
of phase space where the smectic phase is strictly stable is highly restricted,
at finite temperature and for weak coupling,
 we expect a much larger region of phase space 
whose behavior is governed by the sliding Luttinger liquid 
ground state.

\begin{figure}
\par\columnwidth20.5pc
\hsize\columnwidth\global\linewidth\columnwidth
\displaywidth\columnwidth
\epsfxsize=3.0truein
\centerline{\epsfbox{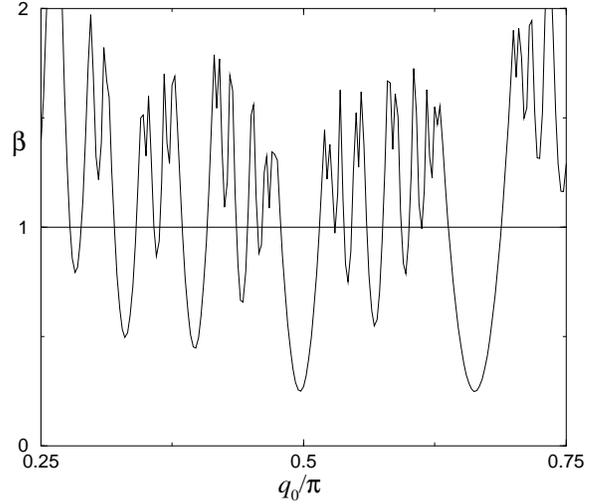}}
\vspace*{.1in}
\caption{ Plot of $\beta \equiv K_{CDW}/K_{SC}$ as a function of
of $q_{0}/\pi$. For $\beta > 1$ there exists a region of $K$ over
which the non-Fermi liquid phase is stable.}
\label{fig:EF1}
\end{figure}

\subsection{The gapless 1:1 array}
    We now consider 1:1 arrays of wires in which
both charge and spin excitations 
are gapless. In this case, there are two Luttinger liquid parameters
$(\kappa_{\rho}, \kappa_{\sigma})$ for the charge and spin modes
respectively, and two velocities $(v_{\rho}, v_{\sigma})$ on each wire.
To maintain gapless Luttinger liquids and SU(2) spin symmetry, we 
do not include any marginal spin-spin coupling terms in the Hamiltonian.
Thus the spin degrees of freedom are represented by the fixed-point
action
\begin{equation} 
 {\it S}_{\phi,\sigma}=\kappa_{\sigma}\sum_{j}\int dx d\tau
\left[v_{\sigma}(\partial_{x} \phi_{\sigma,j})^{2} + {{(\partial_{\tau}
\phi_{\sigma,j})^{2}}\over {v_{\sigma}}}\right]
\end{equation}
with $\kappa_{\sigma}=1$. In a more general treatment, one could
include spin-spin coupling terms and consider their relevance/irrelevance,
maintaining, however, the SU(2) symmetry of the spin-sector. We leave
that for future consideration. The charge modes are still represented
by Eq. (2.9), with $\kappa(q_{\perp})$ and $v(q_{\perp})$ replaced
by $\kappa_{\rho}(q_{\perp})$ and $v_{\rho}(q_{\perp})$. The form of 
$\kappa_{\rho}$ is still given by Eq.~(\ref{kappa}). 

            We again consider the relevance of single-particle,
CDW and SC tunnelling. The SC and CDW tunneling were already considered
in the previous subsection. When the spin variables are included,
Eqs.~(2.11) and (2.12) become
\begin{eqnarray} 
{S}_{{\rm SC},n} &=& {\mathcal J}_{n} \int dx d\tau\sum_{j}\cos[\sqrt{2 \pi}
(\theta_{\rho,j}-{\theta_{\rho,j+n}})] \nonumber \\
& &\cos(\sqrt{2 \pi}\phi_{\sigma,j})
\cos(\sqrt{2 \pi}\phi_{\sigma,j+n}) \nonumber \\
{S}_{{\rm CDW},n} &=& {\mathcal V}_{n} \int dx d\tau \sum_{j}
\cos[\sqrt{2 \pi}
(\phi_{\rho,j}-{\phi_{\rho,j+n}})] \nonumber \\
& &\cos(\sqrt{2 \pi}\phi_{\sigma,j})
\cos(\sqrt{2 \pi}\phi_{\sigma,j+n}).
\label{sccdw2}        
\end{eqnarray}
The $\phi_{\sigma}$ variables now contribute to the 
dimensions of these terms. Because $\kappa_{\sigma}$
is constrained to be 1, the contribution
of the $\sigma$ variables is trivial, and
the dimensions of these terms are  given by 
\begin{eqnarray}
\Delta_{{\rm SC},n}&=&\Delta_{{\rm SC},n}^{\rm (gap)} + 1 \nonumber \\
\Delta_{{\rm CDW},n}&=&\Delta_{{\rm CDW},n}^{\rm (gap)} + 1,
\label{Deltas2}
\end{eqnarray}
where $\Delta_{{\rm CDW},n}^{\rm (gap)}$ and 
$\Delta_{{\rm SC},n}^{\rm (gap)}$ are given by Eqs .~(2.13)
and (2.14) with $\kappa$ replaced by $\kappa_{\rho}$. 

   Since the $\sigma$ variables are no longer gapped,
single-electron tunneling is no longer irrelevant.
Single-particle hopping is described by operators such as 
${R}_{j, \uparrow}{R}^{\dagger}
_{j+n,\uparrow}$, which can be represented by terms of the form
\begin{eqnarray}
{\it S}_{{\rm el},n}=&\int& dx d\tau \sum_{j}{\mathcal T}_{n} e^{-i \sqrt{\pi \over 2}
(\phi_{\rho,j}-\phi_{\rho,j+n})} e^{-i \sqrt{\pi \over 2}
(\theta_{\rho,j}-\theta_{\rho,j+n})} \nonumber \\
& \times&\left\{e^{-i\sqrt{\pi \over 2}(\phi_{\sigma,j}-\phi_{\sigma,j+n}
+\theta_{\sigma,j}-\theta_{\sigma,j+n})}\right\}.
\label{singleelectron}
\end{eqnarray}
 The expectation value of the term in the curly bracket
goes as $L^{-1/2}$ as system size $L$ goes to infinity.
Thus $\langle S_{\rm el} \rangle \sim L^{2 - \Delta_{{\rm el},n}},$ 
where
\begin{equation}
\Delta_{{\rm el},n}= {1 \over 4}[\Delta_{{\rm CDW},n}^{\rm (gap)}
+ \Delta_{{\rm SC},n}^{\rm (gap)}] + {1 \over 2}.
\label{elDelta}
\end{equation}
Regions of phase space where ${S}_{{\rm SC},n}$
is relevant corresponds to the superconducting phase, 
whereas regions of
relevance of ${S}_{{\rm CDW},n}$
correspond to the CDW crystal phase. Regions where
 both these are irrelevant, but single-particle hopping is
relevant correspond to the Fermi-metal phase. For a stable
smectic metal phase, we require that all these operators be
irrelevant. The superconducting and 
CDW coupling terms are irrelevant if
\begin{equation}
   \Delta_{{\rm SC},n}^{\rm(gap)} > 1, \ \ \ 
\Delta_{{\rm CDW},n^{'}}^{\rm (gap)}>1,
\end{equation}
for all $n$ and $n^{'}$
The condition for single particle hopping to be irrelevant
is that
\begin{equation}
   \Delta_{{\rm SC},n}^{\rm (gap)} + \Delta_{{\rm CDW},n}^{\rm (gap)} > 6
\end{equation}
for all $n$.

     We now proceed exactly as for the gapped case, assuming $\kappa_{\rho}$
to have a form as given by Eq.~(\ref{kappa}). As before, we may
write $\Delta_{{\rm CDW},n}^{\rm (gap)}=a_{n}/K$, $\Delta_{{\rm SC},n}
^{\rm (gap)}=b_{n}K$, 
$K_{\mathrm{SC}}={\mathrm{max}}_{n} (1/a_{n})$ and  
$K_{\mathrm{CDW}}={\mathrm
{min}}_{n} (b_{n})$. 
Provided $K_{\rm CDW}/K_{\rm SC} > 1$, there is a window of $K$,
$K_{SC}<K<K_{CDW}$ where the system is stable with respect to 
both the CDW and superconducting couplings. For
the smectic metal phase to be stable, the single particle
hopping has to be irrelevant as well, which indicates
that
\begin{equation}
  {a_{n} \over K} + b_{n}K > 6.
\end{equation}
This condition is violated for $K$ lying between $K_{-}$
and $K_{+}$ where $K_{-}={\rm min}_{n} K_{-,n}$
and $K_{+}={\rm max}_{n} K_{+,n}$, with
\begin{equation}
K_{\pm,n}= {{3 \pm \sqrt{9-a_{n}b_{n}}} \over b_{n}}.
\end{equation}
The single electron hopping is relevant in a large
region of phase space, indicating an instability towards a Fermi 
liquid (FL) phase. We
     write $\lambda_{1}$ and $\lambda_{2}$ as functions of 
$\Delta$ and $q_{0}$, (see Eqn.~\ref{lambdas}), and set 
$\Delta=10^{-5}$. Higher order terms involving $\theta_{\sigma}$
and $\phi_{\sigma}$ in general are less relevant in this case,
and we do not need to consider the whole set of operators \cite{footnote2}.
 Depending on $q_{0}$,
we have the following possibilities for phases as $K$
is increased :
\begin{enumerate}
\item Superconductor(SC) $\rightarrow$ Fermi liquid(FL)$\rightarrow$ CDW crystal;
\item SC $\rightarrow$ Smectic Metal(SM) $\rightarrow$ CDW crystal;   
\item SC $\rightarrow$ FL $\rightarrow$ SM $\rightarrow$ CDW crystal;   
\item SC $\rightarrow$ SM $\rightarrow$ FL $\rightarrow$ SM $\rightarrow$ 
CDW crystal.
\end{enumerate}
The phase diagram is complicated, and
we plot  a region of
$K$, $q_{0}$ space in figure 4, in the absence of disorder.
Backscattering due to
disorder  is irrelevant for $\Delta^{\rm(gap)}_{{\rm CDW},\infty}>2$,
which is automatically satisfied in the SM phase.
\begin{figure}
\par\columnwidth20.5pc
\hsize\columnwidth\global\linewidth\columnwidth
\displaywidth\columnwidth
\epsfxsize=3.0truein
\centerline{\epsfbox{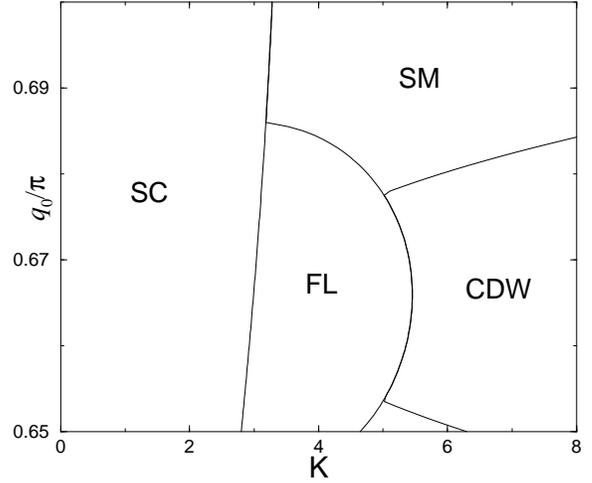}}
\vspace*{.1in}
\caption{A plot of the phase diagram in $q_{0}$, $K$ space
with $\Delta=10^{-5}$. SC stands for Superconducting,
FL for Fermi liquid, SM for smectic metal, and 
CDW for charge density wave crystal.} 
\label{fig:phase}
\end{figure}

\subsection{The three-dimensional anisotropic sliding phase}

 We now turn to three-dimensional 2:1 arrays with wires on a
periodic 2D lattice with primitive translation vectors 
${\vec {a}_{1}}$ and ${\vec{a}_{2}}$. Each wire occupies a position 
$n_{1} {\vec a}_{1} + n_{2} {\vec{a}}_{2}$ on a 2D lattice
and is labeled by the integer valued vector ${\bf { n}}
= (n_{1},n_{2})$.
We will focus on the spin-gapped case, though extensions
to the gapless case proceed exactly as in the previous subsection. 
Taking into account inter-wire density-density and
current-current interactions we could write down an
action of the form (2.7), but where we now have sums
over columns in a 2D lattice, with $j_{\mu,{\bf n}}(x,\tau)
\rightarrow j_{\mu,{\bf n}}(x,\tau)$ and $ {\bf n}=(n_{x},n_{y})$.
When transformed to Fourier space, this action becomes 
\begin{eqnarray}
    S & = &{1 \over 2} \int {{d^{4}Q}\over{(2 \pi)^{4}}}
   [V^{\theta}({\vec q}_{\perp}) q_{\parallel}^{2} |\theta|^{2} \nonumber \\
 & & + V^{\phi}({\vec q}_{\perp} ) q_{\parallel}^{2}|\phi|^{2}
 -i \omega q_{\parallel}\{ \theta^{*} \phi_{x} + {\mathrm{c.c.}}\}
\label{threeaction}
\end{eqnarray}	
where ${\bf Q}=(\omega, q_{\parallel}, {\vec q}_{\perp})$
with ${\vec{q}}_{\perp}$  a vector in the 
first Brillouin zone of the 2D  lattice
of columns. We choose
the $x$-axis to lie along wires, so that ${\vec q}_{\perp}
= (q_{y},q_{z})$.  The $\theta$ or
the $\phi$ variables may be integrated out, giving us the effective actions
\begin{eqnarray}
S_{\theta} &=&\int 
{{d^{4}Q}\over{(2 \pi)^{4}}} {1 \over 2}
\kappa({\vec q}_{\perp})\left\{ {1 \over v({\vec q}_{\perp})}
\omega^{2} + v({\vec q}_{\perp})q_{\parallel}^{2}\right\}  |\phi({\bf Q})|^{2}
\nonumber \\
S_{\phi}&=& \int {{d^{4}Q}\over{(2 \pi)^{2}}}
{1 \over 2\kappa({\vec q}_{\perp})}
\left\{ {1 \over v({\vec q}_{\perp})}
\omega^{2} + v({\vec q}_{\perp})q_{\parallel}^{2}\right\} 
|\theta({\bf Q})|^{2}
\end{eqnarray}
where
\begin{eqnarray}
\kappa({\vec q}_{\perp})&=&\sqrt{V^{\phi}({\vec q}_{\perp})/ 
V^{\theta}({\vec q}_{\perp})} \nonumber \\
v({\vec q}_{\perp})&=&\sqrt{V^{\phi}({\vec q}_{\perp})
V^{\theta}({\vec q}_{\perp})}.
\end{eqnarray}
In three
dimensions it turns out that stability of the sliding
phase with respect to the complete set of operators requires
an even further fine-tuning of the generalized current-current
coupling terms. In particular, $\kappa(q_{y},q_{z})$ should
have a minimum $K \Delta$ at some $q_{y}=q_{0,y}$, $q_{z}=q_{0,z}$, with
both $\Delta$ and the second derivative of $\kappa/K$ being
much smaller than unity at the minimum.
Let us consider two examples of the form
that $\kappa(q_{y},q_{z})$ could assume in order to 
obtain a stable sliding phase.

\begin{figure}
\par\columnwidth20.5pc
\hsize\columnwidth\global\linewidth\columnwidth
\displaywidth\columnwidth
\epsfxsize=2.5truein
\centerline{\epsfbox{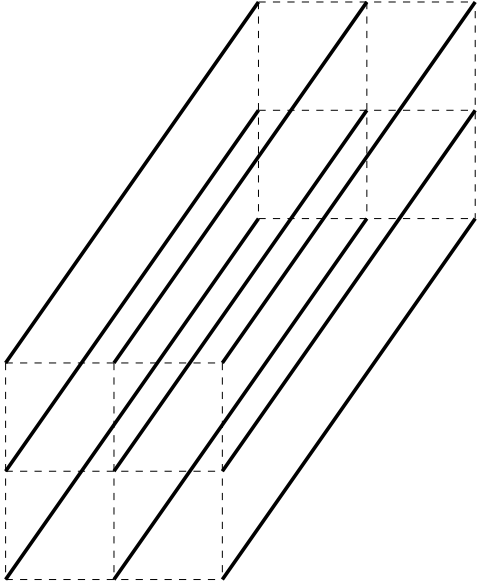}}
\vspace*{.1in}
\caption{A 3-dimensional array of quantum wires.} 
\label{fig:wires2}
\end{figure}
    
   The first example is one which is symmetric with respect
to $q_{y}$ and $q_{z}$. We assume that the wires
are arranged in a square or rectangular 
pattern, and align the $y$- and $z$- axes along the
edges of the rectangle. We consider tha form:
\begin{eqnarray}
\kappa(q_{y},q_{z})&=&K [1 + \lambda_{1} \cos(q_{y}) + \lambda_{1} \cos(q_{z})
\nonumber \\
& &+ \lambda_{2} \cos(q_{y})\cos(q_{z})]^{2}.
\end{eqnarray}
$\lambda_{1}$ and $\lambda_{2}$ are adjusted such that $\kappa$
has a minimum $K \Delta^{2}$ at $q_{y}=q_{z}=q_{0}$.
This gives
\begin{eqnarray}
\lambda_{1}&=& -{(1 - \Delta) \over \cos(q_{0})},
\nonumber \\ 
\lambda_{2}&=& {{(1 - \Delta)} \over {\cos^{2}(q_{0})}}.
\end{eqnarray}
Close to $(q_{y},q_{z})=(q_{0},q_{0})$, we can expand $\kappa$ as
\begin{equation}
\kappa \simeq K [\Delta + \lambda_{2}
(q_{y}-q_{0})(q_{z} -q_{0})]^{2} .
\end{equation}
As before $\Delta_{{\rm SC},{ {\bf m}}}=K a_{{\bf m}}$ and
 $\Delta_{{\rm CDW},{{\bf n}}}= K b_{{\bf n}}$, where 
${{\bf m}}$ and ${{\bf n}}$ are now vectors.
The sliding phase is stable provided
\begin{equation}
\beta \equiv {K_{\mathrm{CDW}} \over K_{\mathrm {SC}}}=
 \left. {{a_{{\bf m}} b_{{\bf n}}} \over 4} \right|_
{{\mathrm{min. wrt.}} {\bf m} \& {\bf n}}> 1.
\end{equation} 
For $\Delta = 10^{-2}$ (or smaller) there 
is a large range of $q_{0}$ where the sliding phase is stable.

  One could also consider the highly anisotropic form 
\begin{eqnarray}
    \kappa(q_{y},q_{z})&=& K[(1 + \lambda_{1} \cos(q_{y})
 + \lambda_{2} \cos(2 q_{y}))^{2} \nonumber \\
& &+ \lambda_{3} (1+\cos(q_{z}))].
\end{eqnarray}
Again, for any $\lambda_{3}$, one can adjust $\lambda_{1}$
and $\lambda_{2}$ to produce a stable sliding phase. 
We conclude by noting that in three
dimensions obtaining a sliding phase requires an even finer
adjustment of parameters than in 2D. At finite temperature,
as before, the region of phase space controlled by
the smectic metal fixed point is expected to expand considerably.

\section{Crossed Sliding Luttinger Liquid phase} 
         Having established regions of stability of the sliding metal phases 
formed from arrays of quantum wires, we now turn to the investigation
of sliding phases formed from crossed arrays of wires.
We consider two basic configurations:
one a two-dimensional system formed from two coupled
2D sliding phases (1:1 arrays) oriented at right angles to each other
and the other a 3D system formed by stacking the crossed two-dimensional
system. The latter three-dimensional system can be constructed 
from two inter-penetrating 3D anisotropic sliding phases
(2:1 arrays), of the type discussed in Sec~2.D, oriented at
right angles to each other. Both the 2D and 3D systems have $C_{4v}$
symmetry. As a result,  their in-plane
conductivities at finite temperature are isotropic at 
long-wavelengths. In this section we demonstrate the
existence of a sliding phase in the crossed arrays
that is stable if the sliding phase in the constituent
arrays is stable. 
The correlation functions in
this phase exhibit power-law decay along the planes, 
and the electric conductivity 
diverges as a power-law in temperature $T$ as $T \rightarrow 0$.

\subsection{Crossed 2D sliding phase}
            We consider now a square grid of wires, starting again 
with the spin-gapped case.  The system consists of 
two  arrays of quantum wires, the $X$- and $Y$- arrays running,
respectively, parallel to the $x$- and $y$-directions. 
Each wire sees a periodic one-electron
potential from the array of wires crossing it.
For simplicity we assume that this periodicity
is commensurate with  bands in the wire. 
This  leads to a new band structure with new band
gaps. It is assumed that the Fermi surface is between gaps so that the
wires would be conductors in the absence of further interactions. By
removing degrees of freedom with wavelengths smaller than the inverse wire
separation, we obtain a new effective theory whose form is identical to
the theory before the periodic potential was introduced. Thus, in the
absence of two-particle interactions between crossed arrays, the system
could be in a phase consisting of two crossed, non-interacting 
smectic-metal states.

 We will now demonstrate the existence of stable sliding phase 
in the crossed arrays. In addition to the interwire couplings
within each array, we need to consider Coulomb interactions between 
wires on the X-array and wires on the Y-array. These 
inter-array couplings are marginal and should be included
in the fixed point. They do not, however, change the
dimensions of the operators, except by renormalizing
$\kappa(q_{\perp})$. For a stable sliding phase, 
additional interactions between the two arrays, such as 
the Josephson and CDW couplings, have to be irrelevant.
We will show that it is possible to tune $\kappa(q_{\perp})$
such that this is indeed the case.

\begin{figure}
\par\columnwidth20.5pc
\hsize\columnwidth\global\linewidth\columnwidth
\displaywidth\columnwidth
\epsfxsize=3truein
\centerline{\epsfbox{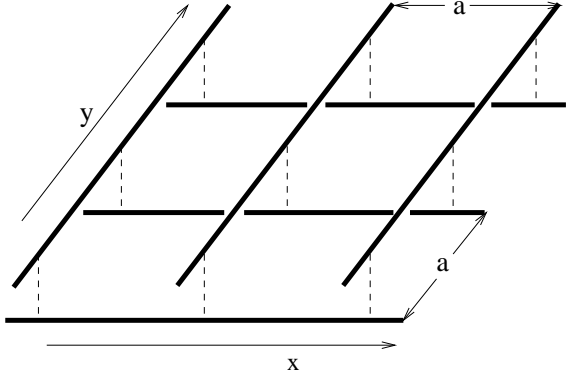}}
\vspace*{.1in}
\caption{Schematic depiction of a 2-dimensional crossed array.} 
\label{fig:wires3}
\end{figure}

The Coulomb interactions between electrons on
intersecting wires give rise to a term in the Hamiltonian
of the form $V^{c}_{m,n}(x,y)\rho_{x,m}(x)\rho_{y,n}(y)$, 
where $\rho_{x,m}(x)$ [$\rho_{y,m}(y)$] is the electron density on the
$m$th wire on the $X$($Y$)-array at position
$x$ ($y$). We expect $V^{c}_{m,n}(x,y)$ to have the form $V^{c}(x - na,
y - ma)$, where $a$ is the distance between parallel wires. 
Thus we represent the interaction between 
the $X$ and $Y$-array as
$$ \int dx dy  [\partial_{x} \phi_{x,m}
V^{c}(x-na, y-mb) \partial_{y} \phi_{y,n}]. $$
If all parameters for the $X$ and $Y$-arrays
are the same, the crossed-grid action as a functional
of the $\theta$ and $\phi$ variables can be written as
\begin{eqnarray}
    S & = &{1 \over 2} \int {{d\omega dq_{x} dq_{y}}\over{(2 \pi)^{3}}}
   [V^{\theta}(q_{y}) q_{x}^{2} |\theta_{x}|^{2}
   + V^{\theta}(q_{x}) q_{y}^{2}|\theta_{y}|^{2}
 \nonumber \\
 & & + V^{\phi}(q_y) q_{x}^{2}|\phi_{x}|^{2}
+  V^{\phi}(q_x) q_{y}^{2}|\phi_{y}|^{2} \nonumber \\
& &+ V^{c}(q_{x},q_{y})q_x q_y \{\phi_{x} \phi_{y}^{*}
 + {\mathrm{c.c.}}\} \nonumber \\
& & -i \omega q_{x}\{ \theta_{x}^{*} \phi_{x} + {\mathrm{c.c.}}\}
 -i \omega q_{y}\{ \theta_{y}^{*} \phi_{y} + {\mathrm{c.c.}}\}] 
\label{action}
\end{eqnarray}	
with obvious definitions for $\phi_x = \phi_x(\omega, q_x, q_y), \phi_y,
\theta_x$, and $\theta_y$.
It should be noted that this is an effective theory with
 $-{\pi \over a} < q_{x},q_{y} < {\pi \over a}$. 
Integrating out the $\phi$ variables, we are left with an effective 
action, which is conveniently expressed in matrix form as
\begin{equation}
  S_{\theta} = {1 \over 2} \int d^{2}k d\omega {\mathbf \theta}_{a} ({\mathbf G}^{-1})_{ab}
{\mathbf \theta}^{*}_{b}
\label{thetaaction}
\end{equation}
where $a=x,y$; $b=x,y$. Here
\begin{equation}
{\mathbf G}^{-1}= \left( \begin{array}{cc}
     {1 \over \kappa_{x}} \left({\omega^{2} \over v_{x}} + 
        v_{x}q_{x}^{2}\right)  &  - V^{c}_{R} \omega^{2} \\
	  - V^{c}_{R} \omega^{2} & 
 {1\over \kappa_{y}}\left({\omega^{2} \over v_{y}} + v_{y}q_{y}^{2}\right)
                      \end{array} \right),
\label{inversegreen}
\end{equation}
where
\begin{eqnarray}
\kappa_{x}({\mathbf q})&=&\sqrt{{\gamma({\mathbf q})} \over
{V^{\phi}(q_{x})V^{\theta}(q_{y})}}, \nonumber \\
v_{x}({\mathbf{q}})&=&\sqrt{{V^{\theta}(q_{y})\gamma({\mathbf{q}})}
\over{V^{\phi}(q_{x})}}, \nonumber \\ 
V^{c}_{R}(\mathbf q) &=& {V^{c}(\mathbf q) \over \gamma(\mathbf q)}\nonumber \\
\gamma({\mathbf{q}})
&=& V^{\phi}(q_{x})V^{\phi}(q_{y}) - (V^{c}({\mathbf q}))^{2},
\label{kappas}
\end{eqnarray}
and $\kappa_{y}({\mathbf q})=\kappa_{x}(P{\mathbf q})$,  
$v_{y}({\mathbf q})=v_{x}(P{\mathbf q})$ where 
$P {\mathbf{q}}=P(q_{x},q_{y})=(q_{y},q_{x})$. 
From Eq.~(3.3) for ${\mathbf G}^{-1}$, we can
calculate
\begin{equation}
{\mathbf G} = {1 \over {\mathcal D}} \left( \begin{array}{cc}
 {1 \over \kappa_{y}} \left({\omega^{2} \over v_{y}} + 
        v_{y}q_{y}^{2}\right)  &   V^{c}_{R} \omega^{2} \\
	   V^{c}_{R} \omega^{2} & 
 {1\over \kappa_{x}}\left({\omega^{2} \over v_{x}} + v_{x}q_{x}^{2}\right)
                      \end{array} \right),
\label{greensfn}
\end{equation}
where
\begin{equation}
  {\mathcal D}={1 \over {\kappa_{x} \kappa_{y}}}
\left({\omega^{2}\over v_{x}} + v_{x}q_{x}^{2}\right) 
  \left({\omega^{2}\over v_{y}} + v_{y}q_{y}^{2}\right)
 - (V^{c}_{R})^{2}\omega^{4} 
\label{discriminant}
\end{equation}
is the determinant of ${\mathbf G}^{-1}$.

In order to determine the dimensions of operators, we calculate the 
leading dependence of correlation functions such $\langle\theta_{x}^{2}
({\mathbf r}, t)\rangle$ on system size $L$. 
Thus we can consider the function
\begin{equation}
 \langle\theta_{x}^{2}({\mathbf r}, t)\rangle=\int{dq_{x}dq_{y}d\omega \over
{(2 \pi)^{3}}}{{\left({\omega^{2} \over v_{y}} + v_{y} q_{y}^{2}\right)}
\over \kappa_{y} {\mathcal D}} 
\label{thetasquared}
\end{equation}
The leading $L$ dependence
is related to  the infrared divergence of the integral
just introduced. This infrared divergence comes purely
from the integration over $q_{x},\omega$. 
We can write the integrand as $\kappa_{x}/({\omega^{2} \over v_{x}}
+ v_{x}q_{x}^{2})$ plus a remaining part, the integral over
which is free of infrared singularities
(see Appendix C for details). Thus, it
is easy to see that the leading $L$ dependence goes as
\begin{equation}
\langle \theta_{x}^{2} \rangle \sim \pi \ln(L) \int {dq_{y} \over 2 \pi}
\kappa(q_{y})
\end{equation} 
where $\kappa(q_{y})= \kappa_{x}(0,q_{y})$.  
Notice that this is precisely what we had for a single array of parallel
wires. A similar analysis yields $\langle \theta^{2}_{y}\rangle
= \langle \theta^{2}_{x}\rangle$. Also note that 
cross-correlations of the form  $\theta_{x}$-$\theta_{y}$ are 
finite as $L$ goes to infinity. 

              We also need to consider correlation functions in 
the $\phi$ variables. To do so, we start with the action 
of Eq.~\ref{action},
and integrate out the $\theta$ variables. The effective action as
a functional of the $\phi$ variables is 
\begin{equation}
     S_{\phi}={1 \over 2} \int {d\omega dq_{x} dq_{y} \over 2\pi^{3}}
     \phi_{a}({\mathbf G_{\phi}}^{-1})_{ab} \phi_{b},
\end{equation}
with $a=x,y$; $b=x,y$. Here
\begin{equation}
{\mathbf G_{\phi}}^{-1}=
\left( \begin{array}{cc}
     {\bar \kappa}_{x} \left({\omega^{2} \over {\bar v_{x}}} + 
       {\bar v}_{x}q_{x}^{2}\right)  &   V^{c} q_{x}q_{y} \\
	   V^{c} q_{x}q_{y} & 
 {\bar \kappa}_{y}\left({\omega^{2} \over {\bar v}_{y}} + 
 {\bar v}_{y}q_{y}^{2}\right)
                      \end{array} \right),
\end{equation}
where
\begin{eqnarray}
{\bar \kappa}(q_{y})&=&(V^{\phi}(q_{y}) /V^{\theta}(q_{y}))^{1/2}, 
\nonumber \\
{\bar v}(q_{y})&=&(V^{\phi}(q_{y}) V^{\theta}(q_{y}))^{1/2}.
\end{eqnarray}
Note that ${\bar \kappa}$
is different from $\kappa$ defined for the $\theta$ correlation
functions. From ${\mathbf G}_{\phi}^{-1}$, we calculate
\begin{equation}
{\mathbf G}_{\phi} = {1 \over {\bar{\mathcal D}}} \left( \begin{array}{cc}
 {{\bar \kappa}_{y}} \left({\omega^{2} \over {\bar v}_{y}} + 
      {\bar v}_{y}q_{y}^{2}\right)  &  - V^{c}q_{x}q_{y} \\
	  - V^{c} q_{x} q_{y} & 
 {\bar \kappa}_{x}\left({\omega^{2} \over {\bar v}_{x}} + 
   {\bar v}_{x}q_{x}^{2}\right)
                      \end{array} \right),
\end{equation}
where
\begin{eqnarray}
  {\bar {\mathcal D}}&=& {\bar \kappa}_{x}{\bar \kappa}_{y}
\left({\omega^{2}\over {\bar v}_{x}} + {\bar v}_{x}q_{x}^{2}\right) 
  \left({\omega^{2}\over {\bar v}_{y}} + {\bar v}_{y}q_{y}^{2}\right)
\nonumber \\
 & & - (V^{c})^{2}q_{x}^{2}q_{y}^{2} 
\end{eqnarray}
is the determinant of ${\mathbf G}^{-1}_{\phi}$.
${\mathbf G}_{\phi}$ can be used to calculate dimensions 
of operators involving $\phi$.
For example the expectation value
\begin{equation}
 \langle\phi_{x}^{2}({\mathbf r}, t)\rangle=\int{dq_{x}dq_{y}d\omega \over
{(2 \pi)^{3}}}{{{\bar{\kappa}}_{x} \left({\omega^{2} \over {\bar v}_{y}} 
+ {\bar v}_{y} q_{y}^{2}\right)}
\over {\bar {\mathcal D}}}
\end{equation}
where again we take
${\pi \over L}< |q_{x}|,|q_{y}|,|\omega|$.
In the integral, as before, the infrared divergence comes purely
from the integral over $q_{x},\omega$. Once more, the infrared
divergent part goes as
\begin{equation}
\langle \phi_{x}^{2}\rangle \sim \pi \ln (L) \int {{d q_{y}} \over {2 \pi}}
{1 \over \kappa(q_{y})}
\end{equation}
where $\kappa(q_{y})=\kappa_{x}(0,q_{y})$ is the same function 
appearing in $\langle \theta_{x}^{2} \rangle$, Eq.~(3.8).

Thus, correlation functions
for $\theta_{x}$ and $\theta_{y}$ can be calculated directly
from Eq. (\ref{thetaaction}). $\theta_{x}$-$\theta_{y}$ cross-correlations
are non-singular, whereas, $\theta_{x}$-$\theta_{x}$ and 
$\theta_{y}$-$\theta_{y}$
correlations have singular parts with exactly the same functional
forms as they have in the absence of coupling between layers,
but with the $\kappa(q)$ function in expressions for the scaling 
exponents replaced by
\begin{equation}
 \kappa(q_{\perp})=\kappa_{x}(0,q_{\perp})=\kappa_{y}(q_{\perp},0) .
\label{kappaq}
\end{equation}
The same holds for $\phi$-$\phi$ correlation functions.
Thus correlation functions within a given array
have the same functional form as for $V^{c}=0$ 
but with different definitions of $\kappa$.
Other than renormalizing
$\kappa(q)$, the coupling $V^{c}_{m,n}$ between the two arrays leaves the dimensions
of all operators {\it unchanged}.
This means that it is possible to choose interchain interactions within
the $X$ and $Y$-grids so that these grids form 2D anisotropic
sliding phases even in the presence of the inter-grid
coupling $V^{c}_{m,n}$.
Equations (\ref{thetaaction}) and (\ref{kappaq}) 
define a 2D non-Fermi liquid with scaling properties to
be discussed in the next section.

First, however, we must verify that it is possible to
choose potentials so that this 2D non-Fermi liquid is stable with respect to
perturbations. All pairwise couplings within a given array,
i.e. ${S}^{X}_{{\mathrm SC},n}$, ${S}^{X}_
{{\mathrm CDW},n}$, ${S}^{Y}_{{\mathrm SC},n}$ and 
${S}^{Y}_{{\mathrm CDW},n}$ defined as obvious generalizations 
of Eqs.~(2.19), can be rendered irrelevant by choosing $\kappa(q_{\perp})$ 
as in the case of an individual array. We must 
also consider Josephson and CDW couplings between the two arrays,
which operate at the points of crossing $(x,y) = (na, ma)$ of wire
$m$ in the $X$-array and wire $n$ of the $Y$-array 
respectively. These take the form
\begin{eqnarray}
{S}^{XY}_{\mathrm{SC}}&=& \sum_{m,n} \int d\tau{\mathcal{J}}^{XY} \cos[\sqrt{2
\pi} (\theta_{x,m}(na) - \theta_{y,n}(ma))] \nonumber \\
S^{XY}_{\mathrm{CDW}} &=& \sum_{m,n} \int d\tau {\mathcal{V}}^{XY} \cos[\sqrt{2 \pi}(\phi_{x,m}(na) - \phi_{y,n}(ma)) \nonumber \\
& &+ 2 k_{F}(m a - n a)]. 
\end{eqnarray}
The dimensions of the cosine operators in the integrands are, respectively,
\begin{eqnarray}
\Delta_{\mathrm{SC},\infty} &\equiv& \int^{\pi}_{-\pi} {dq \over 2 \pi} \kappa(q)
= K \nonumber \\
\Delta_{\mathrm{CDW},\infty} &\equiv& \int^{\pi}_{-\pi} {dq \over 2 \pi} {1
\over \kappa(q)} \simeq {1 \over K} {1 \over \sqrt{C \Delta}} ,
\end{eqnarray}
where we assume that $\kappa(q)$ has the form given by Eq. (3.17),
 $\Delta$ is defined as before, and $C \equiv \kappa''(k_{0})/2K$.
If $\kappa$
is chosen such that Eq.\ (\ref{kappaq}) is satisfied for each array, then
${S}^{XY}_{\mathrm{SC}}$ and ${S}^{XY}_{\mathrm{CDW}}$
are automatically irrelevant. Thus, we do not need any further fine tuning
of $\kappa$ to get a stable CSLL phase. 

Having established a region of stability of the CSLL phase,
we now investigate the nature of the correlation functions.
Consider once more the correlation function 
\begin{equation}
G^{X}_{\phi}(x,y=m a) \equiv \langle
e^{i \phi_{x,m}(x,\tau) - i \phi_{x,0}(0,\tau)+ 2  k_{F} x} \rangle 
+ {\rm c.c.},
\end{equation}
 which corresponds
to the component of the density-density correlation function modulated
at $2 k_{F}$. In the absence of terms such 
as $S^{XY}_{\mathrm{CDW}}$, this correlation function vanishes
for $y \neq 0$. Thus, though irrelevant,
the presence of $S^{XY}_{\rm CDW}$ changes the nature of the correlation 
functions. In its presence, to lowest
order in ${\mathcal V}^{XY}$, we obtain
\begin{eqnarray}
G^{X}_{\phi}(x,y)&\approx& {({\mathcal V}^{XY})^{2} \over 4} e^{i 2 k_{F}
(x + y)} 
\int dx_{1}[ \nonumber \\
& &\langle e^{i \sqrt{2 \pi}(\phi_{x}(x,y=ma)-\phi_{x}(x_{1},y=ma))}
\rangle_{0} \nonumber \\
& & \times  \langle e^{i \sqrt{2 \pi}(\phi_{y}(x_{1},y)-\phi_{y}(x_{1},0))}
\rangle_{0} \nonumber \\
& &\times \langle e^{i \sqrt{2 \pi}(\phi_{x}(x_{1},0)-\phi_{x}(0,0))}
\rangle_{0}] + c.c. ,
\label{GXY}
\end{eqnarray}
where $\langle \rangle_{0}$ is the expectation value with
respect to the CSLL fixed point.
We need the asymptotic form of the correlation function
for large $x$, $y$. Following Eq.~(2.12) we obtain, for example,
\begin{equation}
\langle e^{i \sqrt{2 \pi}(\phi_{x}(x,0)-\phi_{x}(0,0))}\rangle_{0}
\approx {A \over {x^{\Delta_{{\rm CDW},\infty}}}}.
\end{equation}
Similarly
\begin{equation}
\langle e^{i \sqrt{2 \pi}(\phi_{y}(x,y)-\phi_{x}(x,0))}\rangle_{0}
\sim {1 \over {y^{\Delta_{{\rm CDW},\infty}}}}.
\end{equation}
Note that in the sliding phase, $\Delta_{{\rm CDW},n}$ is greater than
unity and thus in Eq.~(\ref{GXY}), the largest contribution 
to the integral comes from $x_{1}$ close to $0$ and to $x$.
It thus follows that for large $x$ and $y$, $G_{\phi}$
goes as 
\begin{equation}
    G_{\phi}^{X} \approx {{C \cos[2 k_{F}(x+y)]} 
\over {(x y)^{\Delta_{{\rm CDW},\infty}}}},
\label{G}
\end{equation}
where $C$ is a constant that is proportional to $({\mathcal V}^{XY})^{2}$.
Thus the correlation function $G_{\phi}$ decays as a 
power law in all directions. Notice that 
$G_{\phi}$ is not isotropic but exhibits a square $C_{4v}$
symmetry.

   The stability of the CSLL phase for the gapless case follows
along the same lines. If there are no marginal inter-array
spin-dependent coupling terms, then $\kappa_{\sigma}=1$,
and we define a renormalized $\kappa_{\rho}(q_{\perp})$. 
The stability of the CSLL phase is
identical to the stability of the smectic phase on a 
single array, with a fixed-point action described by 
 the renormalized function $\kappa_{\rho}(q_{\perp})$. 
Also, proceeding as in Eqs.~(\ref{GXY}) to (\ref{G}), we
now expect the single-electron correlation functions to exhibit
power-law decay in all directions.

\subsection{Crossed 2:1 array}
 
  The above analysis can also be extended quite easily to a three-dimensional
stack of alternate 2D $X$- and $Y$-arrays. We could also think of
such a stack as a three-dimension array of wires running along the
$X$-axis, intermeshed with a 3D array of wires running in
the $Y$-direction. Thus the fixed point action would be of
the form $S_{X} + S_{Y} + S_{XY}$ where $S_{X}$ and $S_{Y}$
are the actions
for the 3D arrays formed by wires running along the $X$-axis
and $Y$-axis respectively, $S_{XY}$ represents the inter-array
Coulomb interactions. Thus the fixed-point action is
\begin{eqnarray}
    S & = &{1 \over 2} \int {{d\omega dq_{x} dq_{y} dq_{z}}\over{(2 \pi)^{4}}}
   [V^{\theta}(q_{y},q_{z}) q_{x}^{2} |\theta_{x}|^{2} \nonumber \\
 & & \ \ \ + V^{\theta}(q_{x},q_{z}) q_{y}^{2}|\theta_{y}|^{2} \nonumber \\
 & & + V^{\phi}(q_y,q_{z}) q_{x}^{2}|\phi_{x}|^{2}
+  V^{\phi}(q_x,q_{z}) k_{y}^{2}|\phi_{y}|^{2} \nonumber \\
& &+ \{V^{XY}(q_{x},q_{y},q_{z})q_{x} q_{y} \phi_{x} \phi_{y}^{*}
 + {\mathrm{c.c.}}\} \nonumber \\
& & -i \omega q_{x}\{ \theta_{x}^{*} \phi_{x} + {\mathrm{c.c.}}\}
 -i \omega q_{y}\{ \theta_{y}^{*} \phi_{y} + {\mathrm{c.c.}}\}] 
\label{action2}
\end{eqnarray}	
where $\phi_{x,y}$ and $\theta_{x,y}$ are functions of
$\omega$, $q_x$, $q_y$, and $q_{z}$, and 
$V^{XY}(q_{x},q_{y},q_{z})$ represents the
 interactions between the $X$- and $Y$-arrays.
For interactions only between nearest neighbor layers, we
obtain $V^{XY}(q_{x},q_{y},q_{z})
= V^{c}(q_{x},q_{y})(1 + e^{i q_{z}})$.
Integrating out the $\phi$ variables, we are left with an effective 
action
\begin{eqnarray}
  S_{\theta}& = & {1 \over 2} \int {{d\omega dq_{x} dq_{y}}\over{(2\pi)^{3}}}
\left[{1 \over \kappa_{x}({\mathbf{q}})}\left(v_{x}({\mathbf{q}}) q_{x}^{2}
+ {{\omega^{2}}\over{v_{x}({\mathbf{q}})}}\right)|\theta_{x}|^{2}
\right. \nonumber\\
& &+ {1 \over {\kappa_{y}({\mathbf{q}})}} \left(v_{y}({\mathbf{q}})
q_{y}^{2}+ 
{{\omega^{2}}\over{v_{y}({\mathbf{q}})}} \right)|\theta_{y}|^{2}
\nonumber \\
& &- \left. \{V^{XY}_{R}({\mathbf{q}}) \omega^{2} 
\theta_{x} \theta_{y}^{*}
 + {\mathrm{c.c.}}\}\right], 
\end{eqnarray}
where 
\begin{eqnarray}
\kappa_{x}({\mathbf q})&=&\sqrt{{\gamma({\mathbf q})} \over
{V^{\theta}(q_{y},q_{z})V^{\phi}(q_{x},q_{z})}}, \nonumber \\
v_{x}({\mathbf{q}})&=&\sqrt{{V^{\theta}(q_{y},q_{z})\gamma({\mathbf{q}})}
\over{V^{\phi}(q_{x},q_{z})}}, \nonumber \\ 
V^{XY}_{R}(\mathbf q) &=& {V^{XY}(\mathbf q) \over \gamma(\mathbf q)}\nonumber \\
\gamma({\mathbf{q}})
&=& V^{\phi}(q_{x})V^{\phi}(q_{y}) - |V^{XY}|^{2}
\label{kappas2}
\end{eqnarray}
and $\kappa_{y}({\mathbf q})=\kappa_{x}(P{\mathbf q})$,  
$v_{y}({\mathbf q})=v_{x}(P{\mathbf q})$ where 
$P {\mathbf{q}}=P(q_{x},q_{y},q_{z})=(q_{y},q_{x},q_{z})$. 
Proceeding exactly as in the previous case, we find 
that 
\begin{equation}
\langle \theta_{x}^{2} \rangle \sim \pi \ln(L) \int {dq_{y} dq_{z} \over 
(2 \pi)^{2}} \kappa_{x}(0,q_{y},q_{z})
\end{equation} 
and
\begin{equation}
\langle \phi_{x}^{2} \rangle \sim \pi \ln(L) \int {dq_{y} dq_{z} \over 
(2 \pi)^{2}} {1 \over \kappa_{x}(0,q_{y},q_{z})}.
\end{equation} 
The stability of the three-dimensional crossed stack is precisely
the same as the stability of a three-dimensional stack of
parallel quantum wires with the Luttinger Liquid parameter
$\kappa(q_{y},q_{z})$ set equal to
$\kappa_{x}(0,q_{y},q_{z})$ of the crossed stack. As before,
there are
no additional singularities due to the coupling between the
crossed arrays.

\section{Transport properties}

We now investigate the transport properties of the sliding
Luttinger liquid phases. 
The conductivities
of an array of parallel wires has been considered by Emery 
{\it {et. al}} \cite{efkl}. In a pure system
the conductivity along a wire is infinite. 
 In the presence of impurities,
the resistivity along the
wires vanishes as \cite{lp}:
\begin{equation}
\rho_{\parallel} \sim T^{\alpha_{\parallel}},
\end{equation}
 with
\begin{equation}
\alpha_{\parallel}= \Delta_{{\mathrm{CDW}},\infty} - 2.
\end{equation}
The  conductivity perpendicular to the wires for an array of parallel
wire can be calculated \cite{efkl,giamarchi} using
the Kubo formula, giving us  
\begin{equation}
\sigma_{\perp} \sim T^{\alpha_{\perp}} 
\end{equation}
with (Ref. \cite{footnote3})
$\alpha_{\perp}=2\Delta_{\mathrm{SC}}-3$,
where $\Delta_{\mathrm{SC}}$
is the minimum of $\Delta_{\mathrm{SC},1}$ and $\Delta_{\mathrm{SC},2}$ 
(For details see Appendix D).
The conductance, $\sigma_c$, arising from the Josephson coupling at the
contact between the crossed wires, can be calculated similarly using
the Kubo formula, and satisfies
\begin{equation}
\sigma_{c} \sim T^{\alpha_{c}},
\end{equation}
where $\alpha_{c}=2\Delta_{\mathrm{SC},\infty}-3$. In this section we focus
on the gapped case. In the gapless case, $\rho_{\parallel}$,
$\sigma_{\perp}$ and $\sigma_{c}$ still exhibit 
power-law behavior even though
the major contribution to perpendicular conductivities may
come from single-particle hopping.

\begin{figure}
\par\columnwidth20.5pc
\hsize\columnwidth\global\linewidth\columnwidth
\displaywidth\columnwidth
\epsfxsize=3.0truein
\centerline{\epsfbox{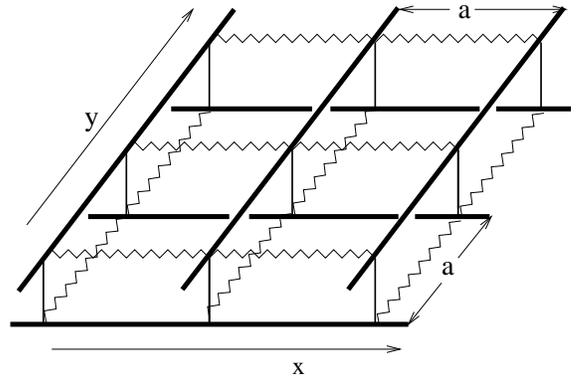}}
\caption{A schematic depiction of the 2D non-Fermi liquid as a resistor
network, with two parallel arrays of wire running along
the $x$ and $y$-axes, with nodes in the $z$ direction}
\label{fig:EF2}
\end{figure}

Thus we can model our 2D non-Fermi liquid as the resistor
network depicted in Fig.~(7) with nodes at the vertical
Josephson junctions between the arrays at $(x,y)=(na, ma)$. The nodes
of the $X$($Y$)-array are connected by nearest neighbor
resistors with conductances $\sigma_{\parallel}=\rho_{\parallel}^{-1}$
if they are parallel to the $x$($y$)-axis and $\sigma_{\perp}$
if they are perpendicular to the $x$-axis($y$-axis). Nearest neighbor
nodes of the $X$ and $Y$-arrays are connected by resistors
of conductance $\sigma_{c}$. In the continuum limit, the 2D
current densities in the plane of the $\alpha$ grids ($\alpha=X,Y$) 	
is $J_{i}^{\alpha}=\sigma^{\alpha}_{ij}E_{j}^{\alpha}$ where
\begin{equation}
\sigma^{X}= \left( \begin{array}{cc}
         \sigma_{\parallel} & 0 \\
   0 & \sigma_{\perp} \end{array} \right),
\end{equation}
\begin{equation}
\sigma^{Y}= \left( \begin{array}{cc}
         \sigma_{\perp} & 0 \\
   0 & \sigma_{\parallel} \end{array} \right)
\end{equation}
and ${\mathbf E}^{\alpha}$ is the 
in-plane electric field in plane $\alpha$. The current per unit area
passing between the planes
is $J_{n}=(\sigma_{c}/a^{2})(V^{X}-V^{Y})$ where $V$ is the
local voltage. In this limit, the local voltages satisfy
\begin{eqnarray}
-\sigma_{ij}^{X}\partial_{i}\partial_{j}V^{X} + {\sigma_{c} \over a^{2}}
(V^{X}-V^{Y}) &=& {\mathcal T}^{X} \nonumber\\ 
-\sigma_{ij}^{Y}\partial_{i}\partial_{j}V^{Y} - {\sigma_{c} \over a^{2}}
(V^{X}-V^{Y}) &=& {\mathcal T}^{Y},
\label{currents}
\end{eqnarray}
where ${\mathcal{T}}^{X}$ and ${\mathcal{T}}^{Y}$ are current
densities (current/area) injected, respectively, into the $X$ and
$Y$-grids. If no currents are injected, then this equation is 
solved by $V^{X}=V^{Y}=-{\mathbf E}\cdot{\mathbf x}$ to produce a
total in-planar current density
\begin{equation}
J_{i}\equiv J^{X}_{i}+J^{Y}_{i}=(\sigma^{X}_{ij} + \sigma^{Y}_{ij})E_{j}
=(\sigma_{\parallel}+\sigma_{\perp})E_{i}.
\end{equation}  
Thus under a uniform electric field, the double layer behaves like 
an isotropic 2D material with in-plane
conductivity $\sigma=\sigma_{\parallel}+\sigma_{\perp}
\simeq \sigma_{\parallel}$, or equivalently with an isotropic
resistivity that vanishes as $\rho_{\parallel} \sim
T^{\alpha_{\parallel}}$. 

      We could also consider currents that
are spatially nonuniform, as they
are, for example, when current is inserted at one point and extracted
from another. In that case,
 there is a crossover from isotropic to anisotropic behavior
at length scale 
\begin{equation}
l= a\sqrt{{\sigma_{\parallel}} \over \sigma_{c}}
\sim T^{-(\alpha_{\parallel}+\alpha_{c})/2}
\label{length}
\end{equation}
that diverges as $T\rightarrow 0$. To illustrate this crossover,
we calculate explicitly
the case where a current $I$ is inserted at a point ${\bf r}_{1}$ on the
X-array and extracted at another point ${\bf r}_{2}$ on the X-array.
Then
\begin{eqnarray}
  {\mathcal T}^{X}&=&I[\delta({\bf r} - {\bf r}_{1}) -
\delta({\bf r} - {\bf r}_{2})], \nonumber \\
{\mathcal T}^{Y}&=& 0.
\label{insertions}
\end{eqnarray}
Using equations (\ref{currents}) and (\ref{insertions}), one can solve
for the resistance between these two points:
\begin{equation}
R={{V^{X}({\bf r}_{1})-V^{X}({\bf r}_{2})} \over I}
=2 \int {{d^{2} q} \over {(2 \pi)^{2}}} {{1 - e^{i {\bf q}\dot({\bf r}_{2}
- {\bf r}_{2})}} \over {g({\bf q})}}
\label{resistance}
\end{equation}
where
\begin{eqnarray}
g({\bf q})&=& {{{\sigma_{c}\over a^{2}}
[(\sigma_{\parallel} + \sigma_{\perp})q_{x}^{2} 
+ (\sigma_{\parallel} + \sigma_{\perp})q_{y}^{2}]}
\over {{\sigma_{c} \over a^{2}} - \sigma_{\perp}q_{x}^{2} - \sigma_{\parallel}
q_{y}^{2}}} \nonumber \\
& &-{{(\sigma_{\perp}q^{2}_{x} + \sigma_{\parallel} q^{2}_{y})
(\sigma_{\parallel}q^{2}_{x} + \sigma_{\perp} q^{2}_{y})}
\over {{\sigma_{c} \over a^{2}} - \sigma_{\perp}q_{x}^{2} - \sigma_{\parallel}
q_{y}^{2}}}.
\label{gq}
\end{eqnarray}
For
\begin{equation}
 \sigma_{c}/a^{2} \gg \sigma_{\parallel} q^{2},
\label{smallq}
\end{equation}
 $g({\bf q})$
takes the simple form $(\sigma_{\parallel} + \sigma_{\perp})q^{2}$.
If $|{\bf r}_{1}-{\bf r}_{2}| \gg l$, with $l$ defined in 
Eq.~(\ref{length}), then the integral over $q$ in Eq.~(\ref{resistance})
is dominated by small $q$ satisfying equation (\ref{smallq}). Thus for
$|{\bf r}_{1}-{\bf r}_{2}| \gg l$, the system has approximately
the same resistance as an {\it isotropic} conductor with 
conductivity $\sigma_{\parallel}$. If
we inserted current at a point on the $X$-array and extracted
it from the $Y$-array we would have the form Eq.~(\ref{resistance}) 
for the resistance with a different function $q^{'}(q)$
whose small $q$ limit is still given by $(\sigma_{\parallel} + 
\sigma_{\perp})q^{2}$. Thus for $|{\bf r}_{1}-{\bf r}_{2}| \gg l$
the resistance is approximately independent of whether
the current is inserted into (or extracted from) the $X$ or
the $Y$-array. More generally,
in a region where ${\mathcal
T}=0$, by inspection of equation~(\ref{currents}) it can be seen
that inhomogeneities in the voltage (or difference between
$V^{X}$ and $V^{Y}$) would heal over the  lengthscale $l$
as defined in Eq.~(\ref{length}), and at longer lengthscales
the system would behave isotropically. This length diverges as
$T \rightarrow 0$ and at $T=0$, current can only be carried
along the wires: the resistance between wires in a grid or between grids is
infinite. 
 
       We could, in addition, investigate the frequency-dependent
zero-temperature conductivity. By arguments similar to those
leading Eqs.~(4.1) to (4.4), we obtain
\begin{eqnarray}
\rho_{\parallel}(\omega) &\sim& \omega^{\alpha_{\parallel}} \nonumber \\
\sigma_{\perp}(\omega) &\sim& \omega^{\alpha_{\perp}} \nonumber \\
\sigma_{c}(\omega) &\sim& \omega^{\alpha_{c}}.
\end{eqnarray}
$\alpha_{\parallel}$, $\alpha_{\perp}$, and $\alpha_{c}$ are the
same as before, though the coefficients are now different (and
complex, in general). At finite $\omega$, the long-wavelength 
resistivity is isotropic as before, and vanishes as
$\rho_{\parallel}(\omega) \sim \omega^{\alpha_{\parallel}}$.

We could also consider extensions of these calculations 
to three-dimensional stacks of crossed arrays.
As we saw in the previous section,
it is possible to get a stable sliding phase in such a system. The conductivity
now has a three-dimensional character, with conductivity along the planes
given by $\sigma_{\parallel}/d$, $d$ being the separation between
adjacent $X$-arrays, but conductivity in the third direction given by 
$\sigma_{c}/a$. Thus the conductivity along the planes
is much larger than the perpendicular conductivity. 

\section{Conclusion}
    In conclusion, we have demonstrated the existence
of non-Fermi metallic phases in two and three dimensions,
that are stable with respect to a wide class of perturbations.
We consider both spin-gapped systems and gapless systems which exhibit
spin-charge separation.  Our central results pertain to the
stability and properties of the crossed sliding Luttinger 
Liquid (CSLL) phase. This 
is a remarkable phase, which
could be identified as a two-dimensional Luttinger liquid. 
The correlation functions in
this phase exhibit power-law decay along the planes, 
and the finite-temperature long-wavelength electric conductivity,
which is  isotropic along the planes,
diverges as a power-law in temperature $T$ as $T \rightarrow 0$. 
The importance of this work is that it provides a perturbative
access to non-Fermi liquid fixed points in two and three
dimensional systems, something that has
proven to be quite difficult in the past \cite{randeria}. This work
could be of significant relevance for higher dimensional
strongly correlated electron systems in general, and to
the normal conducting phases of the cuprates in particular.

  RM and TCL acknowledge support from the National Science
Foundation under grant DMR00-96531. We 
acknowledge useful discussions with S. Kivelson, 
E. Fradkin and A. Vishwanath.

\appendix
\section{}

   In this appendix we sketch out the steps leading to the
asymptotic form for the correlation function $G_{\phi}(x, \tau=0)$
for large $x$ (see Eqs.~(2.12) and (2.13)). 
\begin{eqnarray}
G_{\phi}(x,0)&=& \langle e^{i \sqrt{2 \pi}(\phi_{j}(x,0) - \phi(0,0))
+ i 2 k_{F} x}\rangle + {\rm c.c.} \nonumber \\
   &=& e^{- 2 \pi \langle (\phi_{j}(x) - \phi_{j}(0))^{2} \rangle}
   \cos(2 k_F x).  
\end{eqnarray}
It can be easily checked that
\begin{eqnarray}
\langle (\phi_{j}(x) - \phi_{j}(0))^{2} \rangle &=&
2 \int {{dq_{\parallel} dq_{\perp} d\omega} \over {(2 \pi)^{3}}}
\nonumber \\
{{1 - \cos(x q_{\parallel})} \over {\kappa(q_{\perp})( v(q_{\perp}) 
q_{\parallel}^{2} + {\omega^{2} \over v(q_{\perp})})}}.
\end{eqnarray}
Next we carry out the integration over $q_{\parallel}$ and
$\omega$ obtaining, for large $x$,
\begin{equation}
\int dq_{\parallel} d\omega {{1 - \cos(x q_{\parallel})} \over {\kappa(q_{\perp})( v(q_{\perp}) 
q_{\parallel}^{2} + {\omega^{2} \over v(q_{\perp})})}}
\simeq {{\pi \log(x)} \over \kappa(q_{\perp})} + F(q_{\perp})
\end{equation}
where $F(q_{\perp})$ is some function of $q_{\perp}$ which
depends on $\kappa$, $v(q_{\perp})$  and the momentum cutoff.
From Eqs.~(A.1) and (A.3) it follows that
\begin{equation}
\langle (\phi_{j}(x) - \phi_{j}(0))^{2} \rangle \simeq
 \left[\int {{dq_{\perp}} \over {2 \pi}}
{1 \over \kappa(q_{\perp})} \right]\log(x) + {\rm const.} .
\end{equation}
Using this, we obtain
\begin{equation}
 G_{\phi}(x,0) \approx {{A_{1} \cos(2 k_{F} x)}
\over { x^{\Delta_{{\rm CDW},\infty}}}},
\end{equation}  
where
\begin{equation}
\Delta_{{\rm CDW},\infty} = \int_{-\pi}^{\pi} {{dq_{\perp}} \over 2\pi}
{1 \over {\kappa(q_{\perp})}}.
\end{equation}
Eqs.~(2.14) and (2.15) follow along similar lines.
  
\section{}
Here we outline how the integral
\begin{equation}
I_{n}=\int_{-\pi}^{\pi} {{d q}\over 2 \pi} {{(1 - \cos n q)}
\over f(q)},
\end{equation} 
needed to calculate $\Delta_{{\rm CDW},n}$, can be solved exactly.
Here
\begin{equation}
f(q)= 1 + \lambda_{1} \cos(q) +  \lambda_{2} \cos (2 q),
\end{equation}
using expression (\ref{lambdas}) for $\lambda_{1}$
and $\lambda_{2}$. We can rewrite
\begin{equation}
f(q)= 2 \lambda_{2}( \cos q + u^{+})(\cos q + u^{-})
\end{equation}
where
\begin{equation}
u^{\pm}={1 \over 2}\left[{\lambda_{1} \over {2 \lambda_{2}}}
\pm i {\it D}\right],
\end{equation}
and 
\begin{equation}
{\it D}^{2}= -{\lambda_{1}^{2} \over {4 \lambda_{2}^{2}}}
+ {2 \over \lambda_{2}} - 2 = {{ 2 \Delta} \over \lambda_{2}}.
\end{equation}
It is easy to check that
\begin{equation}
I_{n}= - {1 \over {\sqrt{2 \Delta \lambda_{2}}}} {\mathrm Im} J_{n}^{+}
\end{equation}
with
\begin{eqnarray}
J_{n}^{+} &=& \int_{0}^{2 \pi} {dq\over{2 \pi}}{{1 - \cos (nq)}\over
{u^{+} + \cos q}} \nonumber \\
&=& {1 \over {\pi i}}[ J_{1,n} + J_{2,n}] \nonumber \\
J_{1,n}&=& \oint {{d z} \over {1 + 2 u^{+} z + z^{2}}} \nonumber \\
J_{2,n}&=& -{1 \over 2} \oint {{d z} \over z^{n}} {[1 + z^{2n}]
\over {1 + 2 u^{+} z + z^{2}}}
\end{eqnarray} 
where $z=e^{iq}$, and the $z$ integral is over the unit
circle centered about the origin. The integrands have 
poles at $z=(0,z_{0}^{+},z_{0}^{-})$:
\begin{equation}
z_{0}^{\pm}= - u^{+} \pm i \sqrt{1 - (u^{+})^{2}}.
\end{equation}

    Using (A8), it is easy to check that 
$z_{0}^{+}\times z_{0}^{-}=1$.
Thus, either $z_{0}^{+}$ or $z_{0}^{-}$ lies inside
the contour of integration (the unit circle). Using the
method of residues, it is now straightforward to calculate
the integrals. We simply need to sum over the residues
of the poles enclosed within the contour of integration.
In order to express our results, we distinguish two cases: 
\begin{enumerate}
\item $|z_{0}^{+}|<1$. Then
\begin{eqnarray}
J_{n}^{+}&=& {2 \over (z^{+}_{0} - z^{-}_{0})} \left[
  1 - {(1 + (z_{0}^{+})^{2n}) \over { 2 (z_{0}^{+})^{n}}} \right]
\nonumber \\
& & - \sum_{m=0}^{n-1} {1 \over {(z_{0}^{+})^{m+1}(z_{0}^{-})^{n-m}}}
\nonumber \\
&=& {2 \over (z^{+}_{0} - z^{-}_{0})}[1 - (z^{-}_{0})^{p}].
\end{eqnarray}
\item $|z_{0}^{-}|<1$. In this case
\begin{eqnarray}
J_{n}^{+}&=& {2 \over (z^{-}_{0} - z^{+}_{0})} \left[
  1 - {(1 + (z_{0}^{-})^{2n}) \over { 2 (z_{0}^{-})^{n}}} \right]
\nonumber \\
& & - \sum_{m=0}^{n-1} {1 \over {(z_{0}^{+})^{m+1}(z_{0}^{-})^{n-m}}}
\nonumber\\
&=& {2 \over (z^{-}_{0} - z^{+}_{0})}[1 - (z^{+}_{0})^{p}].
\end{eqnarray}
\end{enumerate}

\section{}

      Here we consider the infrared divergence of the integral
\begin{equation}
I=\int{dq_{x}dq_{y}d\omega \over
{(2 \pi)^{3}}}{{\left({\omega^{2} \over v_{y}} + v_{y} q_{y}^{2}\right)}
\over \kappa_{y} {\mathcal D}} 
\end{equation}
where 
\begin{equation}
  {\mathcal D}={1 \over {\kappa_{x} \kappa_{y}}}
\left({\omega^{2}\over v_{x}} + v_{x}q_{x}^{2}\right) 
  \left({\omega^{2}\over v_{y}} + v_{y}q_{y}^{2}\right)
 - (V^{c}_{R})^{2}\omega^{4}. 
\label{discriminant2}
\end{equation}
At first sight, it may be appear that the integral is divergence
free, since by power counting, there are two powers of
$Q$ (where ${\bf Q}=(\omega,q_{x},q_{y})$) in the numerator
multiplying $d^{3}Q$, and four powers of $Q$ in the denominator.
This seems to indicate that the integral is finite as $L \rightarrow \infty$.
Notice, however, that if $V^{c}_{R}$ is set equal to zero in ${\mathcal D}$,
the integral can be written as
\begin{equation}
I= \int {dq_{x}dq_{y}d\omega \over {2 \pi^{3}}} 
{\kappa_{x}(q_{x},q_{y}) \over {{\omega^{2} \over v_{x}} + v_{x} q_{x}^{2}}},
\end{equation} 
which is clearly infrared divergent. This  divergence 
comes purely from the integration over $q_{x}, \omega$. It turns 
out that even in the presence on $V^{c}_{R}$ the divergence
comes purely from the integration  over $q_{x},\omega$.
To obtain the infrared divergent part we write
\begin{equation}
 {{\left({\omega^{2} \over v_{x}} + v_{x} q_{x}^{2}\right)}
\over \kappa_{x} {\mathcal D}} = 
{\kappa_{x}(0,q_{y}) \over {{\omega^{2} \over v_{x}} + v_{x} q_{x}^{2}}}
+ R
\end{equation}
where $R$ is the remaining piece, and our task is to show that its 
integral has no infrared divergence. Let us write 
$R= R_{1} + R_{2}$ where
\begin{eqnarray}
R_{1}&=& {{ -\kappa_{x}(q_{x},q_{y}) - \kappa_{x}(0,q_{y})} \over 
{{\omega^{2} \over v_{y}} + v_{y} q_{y}^{2}}}, \nonumber \\
R_{2}&=& {{\left({\omega^{2} \over v_{y}} + v_{y} q_{y}^{2}\right)}
\over \kappa_{y}} \nonumber \\
 & & \times \left[ {1 \over {\mathcal D}} - {{\kappa_{x} \kappa_{y}}
\over {\left({\omega^{2}\over v_{x}} + v_{x}q_{x}^{2}\right) 
  \left({\omega^{2}\over v_{y}} + v_{y}q_{y}^{2}\right)}} \right].
\end{eqnarray}
The integral of $R_{1}$ has no infrared
divergence. To check that this is true for $R_{2}$ as well, we
note that in the expression for $R_{2}$ the term in the square brackets 
can be written as
\begin{equation}
 {{\kappa_{x} \kappa_{y} V_{R}^{c} \omega^{4}} \over
{{\mathcal D} \left({\omega^{2}\over v_{x}} + v_{x}q_{x}^{2}\right) 
  \left({\omega^{2}\over v_{y}} + v_{y}q_{y}^{2}\right)}}.
\end{equation}
Now, by noticing the powers of $\omega, q_{y}$,
it is easy to see that the integration of $R_{2}$ has no divergence. 
Thus the infrared divergent part of $I$ can be written
as
\begin{equation}
\int{dq_{x}dq_{y}d\omega \over {(2 \pi)^{3}}} 
{\kappa_{x}(0,q_{y}) \over {{\omega^{2} \over v_{x}} + v_{x} q_{x}^{2}}}.
\end{equation}
Equation (3.8) now follows easily.

\section{}
         We demonstrate explicitly the leading dependence of the
perpendicular conductivity on temperature.  According to the
Kubo formula the transverse conductivity is given by
\begin{equation}
       \sigma_{\perp}(\omega)= {i \over \omega}\left[\Pi_{\perp}(
\omega) + {n_{0} e^{2} \over m} \right],
\end{equation}
where the first term represents the paramagnetic contribution with
\begin{eqnarray}
\Pi_{\perp}(\omega) &=& - i\sum_{j}\int dx\int_{-\infty}^{\infty} dt \Theta(t)
e^{i \omega t} \nonumber \\
& & \langle [J_{\perp}(x,j,t),J_{\perp}(0,0,0)]\rangle
\end{eqnarray}
being the retarded current-current correlator,
and the second term represents the diamagnetic contribution.
The step function $\Theta(t)$ may be written as $(1 + {\rm Sign}(t))/2$
where Sign($t$) is $+1$ for positive $t$, and $-1$ for negative $t$.
In the spin-gapped
case the contribution to the paramagnetic part comes  from superconducting
pair-hopping. The paramagnetic and diamagnetic terms
can be combined to give
\begin{eqnarray}
\sigma_{\perp}(\omega)&= &{i \over  \omega}
\left[ \int dx \int dt e^{i \omega t} (1 +  {\rm Sign}(t)) \right.  \nonumber \\
& & \left.\times[\Pi^{>}(x,t) - \Pi^{<}(x,t)] - (\omega = 0)\right]
\end{eqnarray}
where 
\begin{eqnarray} 
\Pi^{>}&=& -i \sum_{j}\langle J(x,j,t) J(0,0,0) \rangle \nonumber \\
\Pi^{<}&=& -i \sum_{j}  \langle J(0,0,0) J(x,j,t) \rangle.
\end{eqnarray}
Since $\Pi^{>}-\Pi^{<}$ is odd in $t$,  the real part of the 
conductivity is given by
\begin{equation}
\sigma_{\perp}^{'}(\omega)= {i \over  \omega}
\left[ \int dx \int dt e^{i \omega t} [\Pi^{>}(x,t) - \Pi^{<}(x,t)] \right]
\label{realcond}
\end{equation}
Note that the DC transverse conductivity is purely real, and
can be obtained from $\sigma_{\perp}^{'}(\omega)$ by taking
the limit $\omega \rightarrow 0$. 
$\Pi^{>}$ is related by analytic continuation to the Matsubara
correlator $\Pi_{M}(x,\tau)$ in the upper half plane of complex $t$
space, and $\Pi^{<}$ is related to $\Pi_{M}(x,\tau)$ in the lower half
plane. Thus we could view the integral in Eq.~(\ref{realcond})
as an integral over the Keldyish contour shown in figure 5a. 
This contour can  be distorted to the contour shown in 
figure (5b). Note that $\Pi^{>}(t + {i \beta \over 2})=
\Pi^{<}(t - {i \beta \over 2})$ where $\beta = {1 \over {k_{B} T}}$ . 
Thus, we obtain
\begin{eqnarray}
\sigma_{\perp}^{'}(\omega)&=& {1 \over  \omega}
 \int dx \int^{\infty}_{-\infty} 
dt \left[e^{i \omega t} \sinh\left({\omega \beta \over 2}\right)\right.
\nonumber \\
& &\left. \Pi^{>}(x,t+{i \beta \over 2}) \right]
\label{realcond1}
\end{eqnarray}

      The next step is to calculate the Matsubara correlator
$\Pi_{M}(x,\tau)$. To begin with, we only consider nearest-neighbor
hoppings between wires. To lowest order in ${\mathcal J}$
and ${\mathcal T}$:
\begin{equation}
\Pi_{M}(x,\tau)= \sum_{j} \langle J_{\perp}(x,j,\tau) J_{\perp}(0,0,0)\rangle,
\end{equation}
where
\begin{eqnarray}
J(x,j,\tau)&=& a {\mathcal J}_{1}[\sin(\theta_{j}(x,\tau)-\theta_{j-1}(x,\tau))
\nonumber \\
& &+ \sin(\theta_{j+1}(x,\tau)-\theta_{j}(x,\tau))].
\end{eqnarray}
Here $a$ is the distance between adjacent wires, and the expectation
value is taken with respect to the sliding fixed-point Lagrangian.
The correlator can be written as:
\begin{eqnarray}
     \Pi_{M}(x,\tau)&=&{\mathcal J}^{2} a^{2}
\langle e^{i [(\theta_{1}(x,\tau)-\theta_{1}(0,0))+(\theta_{0}(x,\tau)
- \theta_{0}(0,0))]} \rangle \nonumber \\ 
&=&{\mathcal J}^{2} a^{2} \exp[- f(x,\tau)]
\end{eqnarray}
where
\begin{eqnarray}
f(x,\tau)&=&\langle
(\theta_{1}(x,\tau)-\theta_{1}(0,0))^{2}  \nonumber \\
& & -(\theta_{1}(x,\tau)-\theta_{1}(0,0))(\theta_{0}(x,\tau)
- \theta_{0}(0,0))\rangle \nonumber \\
&=& {1 \over \beta} \sum_\omega \int dq_{\perp}dq_{x}[\langle
\theta(q_{x},q_{\perp},\omega)\theta^{*}(q_{x},q_{\perp},\omega)\rangle
\nonumber \\
& & \times(1 - \cos(q_{x}x+\omega\tau))\times(1 - \cos q_{\perp})].
\end{eqnarray}


\begin{figure}
\par\columnwidth20.5pc
\hsize\columnwidth\global\linewidth\columnwidth
\displaywidth\columnwidth
\epsfxsize=2.5truein
\centerline{\epsfbox{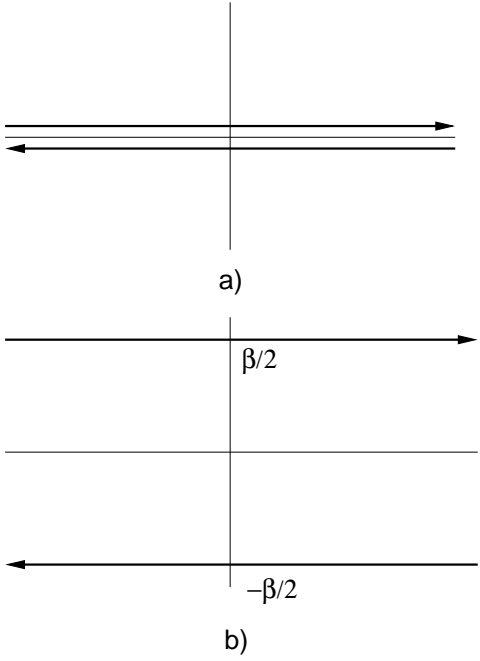}}
\vspace*{.1in}
\caption{a) The Keldyish contour in complex $t$-plane, with
real time along the X-axis. In b) we depict how the contour is 
deformed in order to evaluate the integral. } 
\label{fig:kubo}
\end{figure}
  
Let us first consider a simpler case, where the velocity $v_{s}$ 
has no dependence with $q_{\perp}$. Then
\begin{eqnarray}
&\Pi&_{M}(x,\tau)=a^{2}{\mathcal J}^{2}_{1} \nonumber \\
       &\times& {{(\pi T a_{x}/v)^{2 \eta}} \over {[\sinh(\pi T
(x/v + i \tau))\sinh(\pi T(x/v - i \tau))]^{\eta}}}
\end{eqnarray}
where $\eta=\delta_{{\rm SC},1}$, and $a_{x}$ is the spatial cutoff along
$x$. Thus we may write
\begin{eqnarray}
    \sigma_{\perp}^{'}&=& a_{y}^{2}{\mathcal J}_{1}^{2} {{\sinh\left({\omega \beta\over 2}\right)} \over \omega} (\pi T a_{x}/v)^{2 \eta} (\pi T)^{-2} v
\nonumber \\
& & \times \int d{\tilde x}d{\tilde t}{{e^{i {\tilde t}\left({\omega \over \pi T}\right)
}}\over {[\cosh({\tilde x}+{\tilde t})\cosh({\tilde x} - {\tilde t})]^{\eta}}}
\end{eqnarray}
where ${\tilde x}=\pi x T/v$ and ${\tilde t}= \pi t T$.
By introducing new variables ${\tilde x}+{\tilde t}$ and
${\tilde x} - {\tilde t}$, we  carry out the above integrals,
giving 
\begin{eqnarray}
\sigma_{\perp}^{'}&=&{\mathcal J}_{1}^{2}a_{y}^{2}(v a_{x})^{2\eta} 
{\sinh\left({\omega \over 2T}\right)\over 8 \omega/T} 
\pi v (\pi T)^{2 \eta - 3} \nonumber \\
& &\times {{\Gamma^{2}\left({\eta \over 2} + {i \omega \over 4 \pi T}
\right) \Gamma^{2}\left({\eta \over 2} - {i \omega \over 4 \pi T}\right)}
\over {\Gamma^{2}(\eta)}},
\end{eqnarray}
where $\eta=\Delta_{{\rm SC},1}$.
The $\omega \rightarrow 0$ limit of the above expression yields
\begin{equation}
\sigma_{\perp}^{'}(\omega=0) \sim T^{(2 \Delta_{{\rm SC},1} - 3)}.
\end{equation}

When the velocity $v$ is a function of $q_{\perp}$, the integral can
no longer be solved exactly. However, the leading $T$ dependence of the
conductivity remains unchanged. To check this, we follow the previous
set of steps and arrive at the expression
\begin{eqnarray}
    \sigma_{\perp}^{'}&=& a_{y}^{2}{\mathcal J}_{1}^{2} {{\sinh\left({\omega 
\over 2 T}\right)} \over \omega} (\pi T a_{x})^{2 \eta} (\pi T)^{-2} 
\int d{\tilde x}d{\tilde t}\left[e^{i {\tilde t}\left({\omega \over \pi T}\right)}\right.
\nonumber \\
& &\left. \prod_{q_{\perp}}
{1\over {[v(q_{\perp})\cosh({{\tilde x}\over v}
+{\tilde t})\cosh({{\tilde x}\over v} - {\tilde t})]^{\Delta_{q_{\perp}}}}}\right]
\end{eqnarray}
where ${\tilde x}=\pi x T$, ${\tilde t}= \pi t T$ and
$\sum_{q_{\perp}} \Delta_{q_{\perp}}=\eta=\Delta_{{\rm SC},1}$.
The result can be expressed in  the scaling form
\begin{equation}
\sigma_{\perp}^{'}(\omega,T)=T^{\eta} F(\omega/T).
\end{equation}
In the limit $\omega \rightarrow 0$, the integral is finite
and $T$ independent, implying $F(0)$ is finite.
Thus $\sigma_{\perp}(\omega = 0) \sim T^{\eta}$,
where $\eta = \Delta_{{\rm SC},1}$ for nearest-neighbor
hopping. In exactly the same manner, we can calculate the 
 contribution to
$\sigma_{\perp}^{'}$ from next-nearest neighbor pair-hopping.
It has the same scaling form as before with $\Delta_{{\rm SC},1}$ replaced by
$\Delta_{{\rm SC},2}$.


\begin{thebibliography}{10}
\bibitem{notation} There has been some inconsistency in the
literature regarding the definitions of $\theta$ and $\phi$.
In this paper, we follow the notation of\cite{efkl}, and
define $\theta$ as the phase variable, and $\phi$ as the
conjugate density variable.
\bibitem{footnote1} In these expressions, $\theta_{i}(x)$ 
cooresponds to $\theta_{i}(x)-\theta_{i,0}$, where $\theta_{i,0}$
refers to the uniform part of $\theta$. Hence we exclude the $q_{\parallel}
=0$ modes from the integrals.
\bibitem{footnote2} We should note, however, that the set
of possible interchain  operators is huge and we havent 
explicitly checked for all of them.
\bibitem{footnote3} In this calculation we have ignored the effect
of disorder. At low temperatures, we expect the temperature
dependence of $\sigma_{\perp}$ to be modified due to
forward scattering from impurities. In particular, for very low 
temperatures, $\alpha_{\perp}= 2 \Delta_{\rm SC} - 2$.
\end{thebibliography}
\end{document}